\documentclass[9pt,aps,pra,twocolumn]{revtex4-1}

\usepackage[english]{babel}
\usepackage{dcolumn}
\usepackage{float}
\PassOptionsToPackage{breaklinks}{hyperref}
\usepackage{hyperref}
\usepackage[svgnames]{xcolor}  
\hypersetup{
	colorlinks,
	linkcolor={red!50!black},
	citecolor={blue!50!black},
	urlcolor={blue!80!black}
}
\usepackage{physics}
\usepackage{mathrsfs}
\usepackage{wasysym}  
\usepackage{mathtools} 

\usepackage[T1]{fontenc}
\usepackage[utf8]{inputenc}
\usepackage{calligra}
\usepackage[scr=boondoxo]{mathalfa} 
\usepackage{bm}
\usepackage{bbm}

\usepackage{textcomp}
\usepackage{graphicx} 

\usepackage{verbatim} 
\usepackage{units}

\usepackage{latexsym}
\usepackage{amssymb}
\usepackage{amsfonts}
\usepackage{amsmath}
\usepackage{amsthm}
\usepackage{longtable}
\usepackage{mathtools}
\usepackage{braket}
\usepackage{siunitx}
\usepackage{color}

\definecolor{tud2a}{RGB}{0, 156, 218}
\definecolor{tud9c}{RGB}{185, 15, 34}
\definecolor{green}{RGB}{54 166 59}

\newcommand{\nss}{\negthinspace}

\newcommand{\nls}{\negthickspace}

\renewcommand{\op}[1]{\hat{#1}}
\renewcommand{\vec}[1]{\bm{#1}}

\newcommand{\tir}{\rho}
\newcommand{\tid}{\underline{\delta}}
\newcommand{\tiD}{\tilde{\Delta}}

\newcommand{\Di}{\Delta_1}
\newcommand{\Dii}{\Delta_2}
\newcommand{\Dizero}{\Delta_{1,0}}
\newcommand{\Diizero}{\Delta_{2,0}}
\newcommand{\Oi}{\Omega_1}
\newcommand{\Oii}{\Omega_2}

\newcommand{\Or}{\Omega_\text{R}}
\newcommand{\Ov}{\tilde{\Omega}_\text{R}}
\newcommand{\tOi}{R_1}
\newcommand{\tOii}{R_2}
\newcommand{\oi}{\omega_1}
\newcommand{\oii}{\omega_2}
\newcommand{\vres}{v_\text{R}}

\newcommand{\tvres}{v_{\text{R},0}}

\newcommand{\dvel}{\Delta v_\mathscr{i}}
\newcommand{\vmean}{\bar{v}}
\newcommand{\vi}{v_1}
\newcommand{\vii}{v_2}

\newcommand{\dev}{\nu}
\newcommand{\dvres}{\nu_\text{R}}

\newcommand{\dvi}{\nu_1}

\newcommand{\dur}{d}

\newcommand{\dist}{\beta}

\newcommand{\ca}{$^{40}\text{Ca}^+$}

\newcommand{\uge}{_{ge}}
\newcommand{\ume}{_{me}}
\newcommand{\ueg}{_{eg}}
\newcommand{\uem}{_{em}}

\newcommand{\siggg}{\op{\sigma}_{gg}}
\newcommand{\sigmm}{\op{\sigma}_{mm}}
\newcommand{\sigee}{\op{\sigma}_{ee}}

\newcommand{\sigeg}{\op{\sigma}_{eg}}
\newcommand{\sigem}{\op{\sigma}_{em}}

\newcommand{\Gi}{\Gamma\ueg}
\newcommand{\Gii}{\Gamma\uem}
\newcommand{\omge}{\omega\ueg }
\newcommand{\omme}{\omega\uem}

\newcommand{\fig}{Fig.\,}
\newcommand{\infig}{in \fig}

\newcommand{\spacefortildeB}{\phantom{\Big)}\nls\nss}
\newcommand{\psa}{{(\text{A})}}
\newcommand{\psbb}{{(\text{B})}}
\newcommand{\psb}{{(\text{B}_1)}}
\newcommand{\psc}{{(\text{B}_2)}}
\newcommand{\tpizero}{t_{\pi,0}}

\newcommand{\popm}{\mathscr{m}}
\newcommand{\popmph}{\popm_\text{R}}
\newcommand{\popml}{\popm_\text{sp}}
\newcommand{\popmse}{\popm_{0}}
\newcommand{\popmre}{\popm_{\text{RE}}}
\newcommand{\popmana}{\popm_\text{ana}}
\newcommand{\dpopm}{\Delta v}
\newcommand{\dpopmph}{\Delta v_\text{R}}
\newcommand{\dpopmse}{\Delta v_0}
\newcommand{\dpopmre}{\Delta v_\text{RE}}
\newcommand{\dpopmana}{\Delta v_\text{ana}}

\DeclareMathAlphabet{\mathpzc}{OT1}{pzc}{m}{it}

\newcommand{\rv}{\vec{r}}

\newcommand{\zop}{\op{z}}
\newcommand{\pzop}{\op{p}_z}

\begin{document}
\title{Raman velocity filter as a tool for collinear laser spectroscopy} 

\newcommand{\affT}{Institute for Applied Physics, Technical University of Darmstadt, Germany}
\newcommand{\affN}{Institute for Nuclear Physics, Technical University of Darmstadt, Germany}

\author{A. Neumann}
\email{Antje.Neumann@tu-darmstadt.de}
\affiliation{\affT}

\author{W. Nörtershäuser}
\affiliation{\affN}

\author{R. Walser}
\affiliation{\affT}

\date{\today}

\begin{abstract}
The velocity distribution of a hot ionic beam can be filtered   
with a narrow stimulated Raman process to prepare a colder subensemble, as substantiated in this theoretical analysis. 
Using two counter-propagating, far-detuned lasers, we can 
define a 
$\pi$-pulse for the resonant velocity
to transfer atoms within the linewidth of the Raman resonance
between the ground-states of a $\Lambda$-system. Spontaneous 
emission from the two single-photon resonances, as well as 
the ground-state decoherence induced by laser noise, 
diminishes the efficiency of the filter. From a comprehensive master equation, we obtain conditions for the
optimal frequency pair of the lasers and evaluate  the 
filter performance numerically, as well as analytically.
If we apply this analysis to current \ca\ ion experiments,
we obtain a sensitivity for measuring high ion acceleration voltages on the ppm level or below. 
\end{abstract}

\keywords{Raman spectroscopy, collinear laser spectroscopy, two-photon spectroscopy, high-voltage measurements, ion beam, Calcium}
\maketitle

\section{Introduction} \label{sec:intro}\vspace{-2mm}
"The wonderful blue opalescence of the Mediterranean Sea" is one of the phenomena that Raman attributes to the effect he discovered \cite{Smekal1923,Raman1928,Raman1965}, a century ago. Inelastic two-photon scattering, as we know it today, has found innumerable applications from solid-state spectroscopy and enhanced microscopic imaging \cite{Ferraro2003} to actively cooling atoms with velocity selective coherent population trapping  \cite{arimondo88}. It is our aim to extend the use of Raman transitions for spectroscopy on fast ion beams with several keV kinetic energy. We propose a Raman velocity filter to selectively prepare the population of a metastable state for subsequent precision spectroscopy.
\subsection{Motivation}\vspace{-2mm}
In the context of fast ion and atom beams, collinear laser spectroscopy \cite{Kaufman1976,Wing1976,Schinzler1978} allows us to perform investigations of optical transitions with high resolution and sensitivity. The salient feature is the kinematic compression of the velocity width due to the electrostatic acceleration that reduces the Doppler width of initially hot thermal samples to the typical natural linewidth of allowed optical dipole transitions \cite{Kaufman1976}. This and the fast transport of the ions make it the ideal tool to study short-lived isotopes with lifetimes in the millisecond range that are produced at on-line facilities \cite{Neugart1985,Otten1989}. Such investigations are usually performed to determine nuclear ground state properties such as spins, charge radii and electromagnetic nuclear moments \cite{Otten1989, Blaum2013, Campbell2016, Neugart2017}. Collinear laser spectroscopy has also been used for ultratrace analysis of long-lived isotopes at very low abundance \cite{Wendt1999} and was proposed as a technique to measure high voltages $U$ with very high precision by Doppler velocimetry \cite{Poulsen1988,Gotte2004,Krieger2011}.

While the kinematic compression of collinear laser spectroscopy can produce spectra with resolution close to the limit of the natural linewidth, this is not always the case. Especially if the ions are generated in a plasma, e.g., in an electron-cyclotron resonance source, or in a region with strong electric fields such as a liquid metal ion source, a substantial residual broadening can remain. Moreover, the ground state of the ion is not always the most appropriate initial level for the spectroscopy. An excited metastable level might be better suited for the purpose of an experiment, for example, if the transition from this level can provide atomic hyperfine fields with better accuracy or a higher angular momentum provides the possibility to determine the nuclear spin. In such cases population transfer has already being used for collinear laser spectroscopy \cite{Baczynska2010} but only while the ions are stored in a linear Paul trap filled with buffer gas to accumulate and store the ions. Pulsed lasers address ions of all velocity classes and efficiently excite them into a higher lying state that has a decay branch into the metastable state to be addressed by collinear laser spectroscopy after ejection from the trap. Such a scheme often suffers from various decay paths into a multitude of levels after the resonant excitation. In a Raman transition, population is transferred between two levels without real occupation of a third level with potential leakage into dark states. Only a single experiment has reported a Raman transition in collinear laser spectroscopy so far: In \cite{Dinneen1991} a transition between two hyperfine components of Y$^+$ was induced using a single laser beam which was frequency modulated with an electro-optic modulator.

Here we investigate theoretically the possible use of Raman transitions in collinear laser spectroscopy to selectively transfer ions from the ground state to a higher lying metastable state. Beams from different lasers have to be used to bridge the large excitation energy. We are particularly interested in the usage of the Raman scheme as a velocity filter to prepare ions with a very narrow velocity distribution in an excited state to perform afterward high-resolution collinear laser spectroscopy on this excited population. As a potential application for such a scheme, we have addressed high-voltage measurements using Ca$^+$ ions.
We investigate the influence of interaction time, atomic velocity, laser linewidth, and laser intensity on the excitation efficiency. Our results suggest that Raman transitions can be used with the available laser beams to considerably improve the measurement accuracy with Ca$^+$ ions for high-voltage measurements. This approach will be tested experimentally in the near future and might become the basis for further improvements of laser-based high-voltage measurements, which is of great interest for several applications, e.g., the neutrino mass measurement of the KATRIN experiment \cite{Thummler2009,Rest2019, Aker2019}.

\subsection{Spectroscopic high-voltage measurements}\vspace{-2mm}
\begin{figure}[t]
	\centering
	\includegraphics{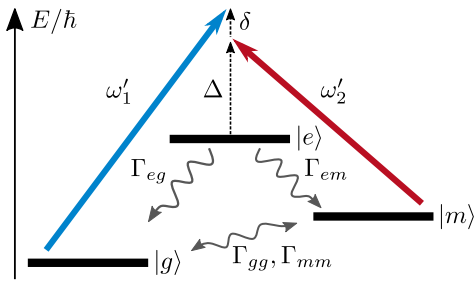}
	\vspace{-2mm}
	\caption{Three-level energy diagram for \ca. Laser 1 induces $eg$-transitions with $\omega_{eg}=\omega_e-\omega_g$ and
		laser 2 couples the $em$-transition with $\omega_{em}=\omega_e-\omega_m$. In the rest frame of the ions, we define a one-photon detuning
		$\Delta=\oii'-\omega_{em}$ and a two-photon detuning $\delta=\oi'-\oii'-\omega_{mg}$ with respect to the Doppler-shifted frequencies $\oi'$, $\oii'$ given by Eq.~\eqref{eq:omdop}.
		The spontaneous decay rates $\Gamma_{eg}$ and $\Gamma_{em}$ couple the excited state $\ket{e}$ to the ground $\ket{g}$ and metastable state $\ket{m}$. Further, laser noise induces ground state decoherence with
		rates $\Gamma_{gg},\Gamma_{mm}$.\vspace{-3mm}}
	\label{fig:3ls}
\end{figure}
Recently, it was demonstrated that an accuracy $s=\Delta U/U$ of at least a few ppm can be reached for high voltages up to \SI{20}{\kilo\volt} \cite{Kramer2018} in laser spectroscopic high-voltage measurements. This is very close to the performance of the world's best high-voltage dividers \cite{Thummler2009,Rest2019}.
In the corresponding measurements two transitions in calcium, shown \infig\ref{fig:3ls}, have been employed:
The $\ket{g=(4s\,^2S_{1/2})} \rightarrow \ket{e=(4p\,^2P_{3/2})}$ resonance transition was first used to transfer population from the
ionic ground level into the metastable level $\ket{m=(3d\,^2D_{5/2})}$ via a sequential stimulated absorption and spontaneous emission cycle.
For this process laser 1 is counter-propagating to the ion beam and the laser frequency determines the longitudinal velocity of the ions required to match the Doppler-shifted resonance condition. Afterwards, the ions are accelerated and the velocity of those ions that are in the metastable state is determined with a second laser tuned to the Doppler-shifted $\ket{m}
\rightarrow  \ket{e}$ transition. The resonance is observed using the fluorescence light emitted in the subsequent decay into the ground state. The frequencies in the laboratory frame of both lasers are measured with a frequency comb and are used to calculate the shift in frequency and the corresponding acceleration voltage.

While an $s=\SI{5}{ppm}$ uncertainty has been achieved by now, we are
investigating other approaches that promise even higher accuracy
for high-voltage measurements. One of the critical issues
using the Ca$^+$ ionic beam is the remaining transverse emittance
of the beam. Due to the 23-MHz width of the resonance transition,
ions with small angles relative to the laser beam direction, might
also be excited and the angle with respect to the laser beam might
be changed during the acceleration with the high-voltage to be measured; however,
measures to avoid this have been taken:
The ion optics of the acceleration region has been designed to suppress such effects by accelerating in the focal region and shaping the beam afterward again into a beam with similar parameters (size and opening angle) as before.
A second point is that several excitations are often needed to transfer the ion from the ground state to the metastable state, which is accompanied by uncontrollable recoil effects due to the momentum transfer in absorption and emission.

\subsection{Raman velocimetry}
\begin{figure}[t]
	\centering
	\includegraphics{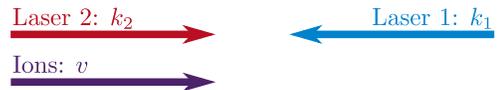}
	\vspace{-3mm}
	\caption{Two counter-propagating lasers with  wave-vectors
		$\vec{k}_1 = -k_1 \vec{e}_z$ and $\vec{k}_2 = k_2 \vec{e}_z$
		interact with the ions, which move with velocity $\vec{v} = v\vec{e}_z$ parallel to laser 2. Wave-numbers and scalar velocities are positive quantities $k_i,v>0$.\vspace{-3mm}}
	\label{fig:laserion}
\end{figure}

\vspace{-2mm}
Here, we elaborate on the possible use of Raman transitions between $\ket{g}$ and $\ket{m}$ by applying a co- and
a counter-propagating laser beam, as depicted \infig\ref{fig:laserion}, to reduce corresponding uncertainties with the existing excitation scheme.
The advantage is that the selectivity of the narrow Raman transition with respect to the atoms initial velocity as well as to the angle between the laser direction and the atoms' movement is considerably higher than for the allowed dipole transitions used so far. This will provide better control of the initial conditions of the atoms prepared in the metastable state before the acceleration.
The theoretical treatment is based on the experimental boundary conditions at the collinear apparatus for laser spectroscopy and applied sciences at TU Darmstadt, where the previous high-voltage measurements were performed. However, the derived models are universally applicable to three-level $\Lambda$-systems together with fast atomic motion.
\subsection{Structure of the article}\vspace{-2mm}
This article is organized as follows. In 
Sec.\,\ref{sec:exp} we present the experimental setup 
and formulate an appropriate  master-equation in the classical kinetic regime in Sec.\,\ref{sec:ww}. Section\,\ref{sec:sol} contains analytical models and their results for the population transfer into the metastable state, matching the numerical results. Therewith we can answer the question if the achieved transferred population into the metastable state is sufficient for the realization of optical high-voltage measurements with spectroscopic precision.

\section{Ion-laser interaction}\label{sec:exp}\vspace{-2mm}
The Raman spectroscopy is formed with two counter-propagating laser beams that interact with \ca, moving with velocity $v$ in the same direction as laser 2, as depicted \infig\ref{fig:laserion}. 
\subsection{Ionic velocity distribution}\vspace{-2mm}
\label{sec:velodist}
In the beam line, ions get accelerated by a high voltage $U$. 
For the  typical value $U=\SI{14}{\kilo\volt}$, one can estimate the mean velocity $\bar{v}$ from energy conservation
\begin{equation}\label{eq:U}
eU = \frac{m\vmean^2}{2} .
\end{equation}
For singly charged \ca, 
one finds $\vmean =\SI{260}{\kilo\meter\per \second}=
\num{8.7e-4}\,c$, which is much smaller than the speed of light $c$ and justifies a non-relativistic treatment.

\begin{figure}[t]
	\hspace{3mm}
	\includegraphics{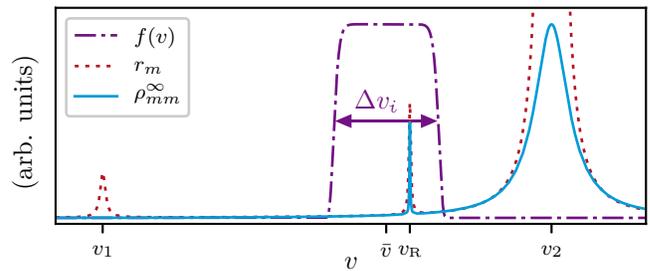}
	\vspace{-6mm}
	\caption{Ionic velocity distribution $f(v)$ vs. velocity $v$ with mean velocity 
		$\vmean$ and width $\dvel$
		(dash-dotted). Superimposed are the initial growth rate $r_m$ \eqref{eq:rate} (dotted) of the metastable state's population from perturbation theory and the exact stationary solution $\rho_{mm}^\infty$ \eqref{eq:laplace} and \eqref{eq:laplacestatsol} (solid), which exhibit resonances at $\vi$, $\vii$ and $\vres$. Note that the velocity distribution emerging from the accelerator is rather flat topped. 
		\label{fig:veldist}}
\end{figure}

Due to technical reasons, the ensemble emerges with an artificial velocity distribution $f(v)$, which is depicted \infig\ref{fig:veldist}. It exhibits an initial residual 
velocity spread $\dvel$=\SIrange{10}{100}{\meter\per \second} (FWHM).
The spectroscopy is performed in an interaction zone of length 
$L=\SI{1.2}{\meter}$.
This gives a mean transit time
$\bar{\tau}=L/\vmean = \SI{4.62}{\micro \second}$. 
Due to the velocity spread, an interaction time spread
$\Delta \tau/\bar{\tau}= \dvel/\bar{v} $ arises. 
For the maximal velocity width 
$\dvel=\SI{100}{\meter\per\second}$, one finds $\Delta \tau=\SI{1.7}{\nano\second}$, which is negligible.
In the short time of the spectroscopy pulse, we 
can neglect binary interactions or other charge effects. 

Due to the large momentum uncertainty $m\dvel\gg \hbar k_i$ compared to the photon momentum recoil, we disregard mechanical light effects. Therefore, the position $z$ and the momentum of the particle $p=mv$ can be treated as parameters. Consequently, observables are obtained by static averaging over the initial phase-space distribution. At first, we will assume that the laser and the ion beam are spatially homogeneous. This is rectified  in 
Sec.~\ref{sec:spatio}, when we consider spatial variations. 

\subsection{Optical Bloch equations}\vspace{-2mm}
\label{sec:ww}
 We assume a closed three-level system for the electronic structure of the ions, consisting of the ground-state manifold $4s\,^2S_{1/2}$, 
the excited state $4p\,^2P_{3/2}$ and the metastable state $3d\,^2D_{5/2}$, depicted \infig\ref{fig:3ls}. The lifetime $\tau_{mg}= \SI{1.168(7)}{\second}$ \cite{Barton2000} of the metastable state is much longer than the duration of the spectroscopy and therefore it is considered as stable. Further calcium data and laser parameters are provided in 
Tables \ref{tab:ionparameters} and  \ref{tab:laserparameters}.

To model the interaction, we use the rest frame of an ion moving with velocity $v$. Thus, the laboratory frame laser frequencies get Doppler-shifted
\begin{align}
\label{eq:omdop}
\omega'_i=\omega_i-\vec{k}_i\vec{v} =
\begin{cases}
 \oi +k_1v,\\
 \oii -k_2v,
\end{cases}
\end{align}
using the vacuum dispersion $\omega_{i}= ck_{i}$. Furthermore, we assume that the ions and lasers propagate exactly in the $z$-direction with no relevant transversal inhomogeneity. This leads to an effective one-dimensional description with the effective
$\Lambda$-Hamilton matrix 
\begin{align}
\label{eq:hamilton}
\mathcal{H}_{ij}=&
\hbar \begin{pmatrix}
 \Delta_1  & \frac{\Omega_1^\ast}{2} & 0 \\
 \frac{\Omega _1}{2} & 0 & \frac{\Omega _2}{2} \\
 0 & \frac{\Omega_2^\ast}{2} & \Delta_2
\end{pmatrix}\!.
\end{align}
It is based on the  electric dipole interaction in the rotating wave approximation
\cite{CohenTannoudji2008} with basis states sorted as $i\in\{g,e,m\}$. The strength of the dipole interaction is measured by the Rabi frequency 
$\Omega_i=-\vec{\epsilon}_i \vec{d}_{ge}\mathcal{E}_i/\hbar$, with dipole matrix element $\vec{d}_{ge}$, laser polarization $\vec{\epsilon}_i$ and electric field amplitude $\mathcal{E}_i$.
The detunings of laser 1 ($\Di$) and laser 2 ($\Dii$) define the one- ($\Delta$) and two-photon detuning ($\delta$), following from the energy diagram \infig\ref{fig:3ls}
\begin{align}
\label{detunings}
\begin{split}
\Di&=  \omega_1+k_1v-\omega_{eg} = \Dizero +k_1 v,\\
\Dii&= \omega_2-k_2v-\omega_{em} = \Diizero -k_2 v\equiv \Delta,\\
\delta &=\Delta_1 - \Delta_2,
\end{split}
\end{align}
denoting  transition frequencies as $\omega_{ij} = \omega_i - \omega_j$.
Herein we will frequently consider the limit of weakly saturated transitions. This is conveniently captured by the  saturation 
parameter $s_i=|\Omega_i|^2/2\Delta_i^2\ll 1$.
From second-order perturbation theory of the Schrödinger equation
$\mathcal{H} \vec{w}_i=\hbar \underline{\Delta}_i\vec{w}_i$ \eqref{eq:hamilton}, one obtains the ac-Stark-shifted eigenfrequencies
\begin{align}
\label{perteng}
\underline{\Delta}_1&= \Delta_1\left(1 +\frac{s_1}{2}\right) ,&
\underline{\Delta}_2&= \Delta_2\left(1  +\frac{s_2}{2}\right),
\end{align}
and
$\underline{\Delta}_3=\Delta_1+\Delta_2-\underline{\Delta}_1-\underline{\Delta}_2$ 
to order $\mathcal{O}(s_1^2,s_2^2)$.

An ensemble of ions interacting with lasers in free space establishes an open quantum system and must be 
described by a master equation
 for the 
semi-classical density operator $\hat{\rho}(t;z,v)$ 
\begin{align}
\label{eq:master}
\begin{split}
\dot{\op{\rho}}=&
-\frac{i}{\hbar}\left[ \mathcal{H},\op{\rho}\right] 
+ (\mathcal{L}_{eg}
+ \mathcal{L}_{em}
+\mathcal{L}_{gg}+
\mathcal{L}_{mm})\op{\rho}
\end{split},
\end{align}
with the Lindblad operators (cf. Appendix~\ref{raman})
\begin{equation}
\mathcal{L}_\lambda\op{\rho}
 \equiv\frac{\Gamma_\lambda}{2}\left(
 2\op{\sigma}_\lambda\op{\rho}\op \sigma^\dagger_\lambda
   -\op \sigma^\dagger_\lambda \op \sigma_\lambda\op{\rho}
   -\op{\rho} \op \sigma^\dagger_\lambda \op \sigma_\lambda\right).
\end{equation}
The first term of the master equation describes the 
coherent dynamics.  
The second and third terms represent spontaneous transitions to the
 ground state $\ket{g}$ and metastable state $\ket{m}$ with decay
rates $\Gamma\ueg$ and $\Gamma\uem$, respectively. 
The forth and fifth term consider ground-state dephasing due to
finite laser linewidths $\Gamma_{gg}$ of laser 1  and 
$\Gamma_{mm}$ of laser 2 
\cite{Dalibard1985,Marksteiner1995,Walser1994,Rosenbluh1998,McIntyre1993,Sturm2014}.  
If one represents the master equation in a basis and 
arranges the matrix elements of $\rho=(\rho_{ij})$ as a list 
accordingly (c.f. Appendix~\ref{app:blochm}), one obtains
\begin{align}
\label{blochL}
\dot{\rho}(t;v)=L(v)\rho(t;v).
\end{align}
Explicitly, these optical Bloch equations (OBEs) read 
\begin{equation}
\label{eq:BEmasterpop}
\begin{split}
\dot\tir_{ee}     &=   -\Gamma\tir _{ee} +\tfrac{i}{2}(\Oi^\ast\tir_{eg}  + 
  \Oii^\ast\tir_{em}-\text{H.c.}),  \\
\dot \tir_{gg}	 &=   \Gi\tir _{ee} + \tfrac{i}{2}(\Oi\tir_{ge} - \Oi^\ast\tir_{eg}), \\
\dot \tir_{mm} &=   
\Gii\tir _{ee}  +
\tfrac{i}{2}(\Oii\tir_{me} - \Oii^\ast\tir_{em})
\end{split}
\end{equation}
for the populations and for the coherences $\rho_{ij}=\rho_{ji}^\ast$
\begin{align}
\label{eq:BEmastercoh}
\dot \tir_{eg}    &=  (i\Delta_1-\Gamma_1)\tir_{eg} + \tfrac{i}{2}[
\Oi(\tir_{ee} - \tir_{gg}) 
-\Oii\tir_{mg} ],\notag\\
\dot \tir_{em}   &=  (i\Delta_2-\Gamma_2)\tir_{em} + \tfrac{i}{2}[\Oii(\tir_{ee} - \tir_{mm}) -
\Oi\tir_{gm}], \notag\\
\dot \tir_{gm}  &=  -(i\delta+\gamma)\tir_{gm} + \tfrac{i}{2}[\Oii\tir_{ge} -\Oi^\ast\tir_{em}],
\end{align}
with composite rates $\Gamma=\Gi+\Gii$, 
$\Gamma_1 = (\Gamma+\Gamma_{gg})/2$, 
$\Gamma_2 = (\Gamma+\Gamma_{mm})/2$, and 
$\gamma=(\Gamma_{gg} +\Gamma_{mm})/2$.

\subsection{Resonance conditions}\vspace{-2mm}
\label{sec:resv}

The objective for using the stimulated Raman transition is to filter a velocity group $\vres$ from the ionic ensemble with a resolution below the natural linewidth. 
From energy conservation (cf. \fig\ref{fig:3ls}) and the ac-Stark-shifted eigenfrequencies $\underline{\Delta}_i$ \eqref{perteng}, one obtains the kinematic condition for the two-photon resonance as 
\begin{equation}\label{eq:tid}
\underline{\delta}(\vres) =\underline{\Delta}_1-\underline{\Delta}_2=0. 
\end{equation}
Thus, the Doppler-shifted laser frequencies must match the 
ac-Stark-shifted transition frequencies of the ground states \eqref{perteng}. This defines the Raman resonance velocity
\begin{align}
\vres
=-\frac{\delta_0}{k_1+k_2}+
\frac{|\Oii|^2-|\Oi|^2}{4 (k_1 +k_2) \tiD}, \label{eq:vres}
\end{align}
where $\tvres=-\delta_0/(k_1+k_2)$ is the dominant contribution and around the Raman resonance we can approximate
\begin{equation}\label{eq:approx}
\frac{\Omega_i^\ast \Omega_j}{\Delta_i(v)} 
\approx\frac{\Omega_i^\ast \Omega_j}{\tiD}, \
\tiD\equiv \Delta_i(\tvres)=\frac{\Diizero k_1+\Dizero  k_2}{k_1+k_2}
\end{equation}
 within the limit of weak saturation.
It is interesting to recognize the magic spot
$|\Omega_1|=|\Omega_2|$, where second order energy shifts cancel in \eqref{eq:vres}.

There are also two rogue resonances at 
velocities $v_1$, $v_2$, 
 where each laser couples resonantly to the excited state
\begin{align}
\label{eq:vres1ph1} 
\underline{\Delta}_1(\vi) &=0,  
&\vi
&=-\Delta_{1,0}/k_1,\\ 
\label{eq:vres1ph2} 
\underline{\Delta}_2(\vii)&=0,  
&\vii
&= \Delta_{2,0}/k_2.
\end{align}

\begin{figure}[t]
	\centering
	\includegraphics{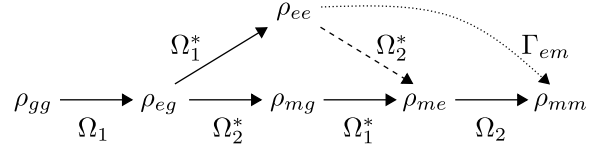}
	\vspace{-3mm}
	\caption{Linkage pattern for two-photon transitions connecting the ground state $\ket{g}$ with the metastable state $\ket{m}$.
	\label{fig:poptransschema}}
\end{figure} 
The width and strength of the resonances are determined by the OBEs \eqref{eq:BEmasterpop} and  \eqref{eq:BEmastercoh}.
From the linkage pattern of \fig\ref{fig:poptransschema}, one
obtains three pathways to reach state $\ket{m}$, starting at $\ket{g}$. 
Perturbatively, the initial growth rate $r_m=\dot{\rho}_{mm}(t=0)$ of the metastable state reads
(c.f. Appendix~\ref{app:blochm})
\begin{align}\label{eq:rate}
r_m=&\frac{\Gamma_1 s_1}{\Gamma}
\left[\Gamma_{em} + 
\Gamma_2 s_2
\left\{1+
\frac{\Gamma\delta}{2(\gamma^2+\delta^2)}
\left(\frac{\Delta_1}{\Gamma_1}-\frac{\Delta_2}{\Gamma_2}\right)
\right.\right.  \notag \\
&+\left.\left.
\frac{\Gamma\gamma}{2(\gamma^2+\delta^2)}
\left(1+ 
\frac{\Delta_1 \Delta_2}{\Gamma_1\Gamma_2 }\right)
\right\}\right],
\end{align}
where we generalize the saturation parameter from the coherent limit to
$s_i=|\Omega_i|^2/2(\Gamma_i^2+\Delta_i^2)$, now broadening the resonances with the linewidths $\Gamma_i$.  
The first two  resonances occur spontaneously at $\Di=0$ and $\Dii=0$,  
while the last two describe the stimulated Raman process at the two-photon resonance $\delta = 0$. Due to laser noise, it acquires the finite linewidth $\gamma$. 
This growth rate $r_m$ is schematically depicted \infig\ref{fig:veldist} together with the stationary solution $\rho_{mm}^{\infty}$ \eqref{eq:laplacestatsol}, also derived in App.\,\ref{app:blochm}. 
The narrow stimulated Raman resonance at $v_R$, is clearly distinguishable from the resonance of laser 1 at $v_1$, where $\ket{e}$ gets populated followed by spontaneous emission into $\ket{m}$.
This process limits the velocity determination, due to the broad tail. Therefore, it is called the rogue resonance in the further course.
In contrast to the \mbox{rate $r_m$,} the stationary solution 
$\rho_{mm}^{\infty}$ suppresses the resonance at $\vii$, because  stimulated emission is compensated with stimulated absorption. 

\subsection{Isolating resonances}
\begin{figure}[t]
	\centering
	\includegraphics{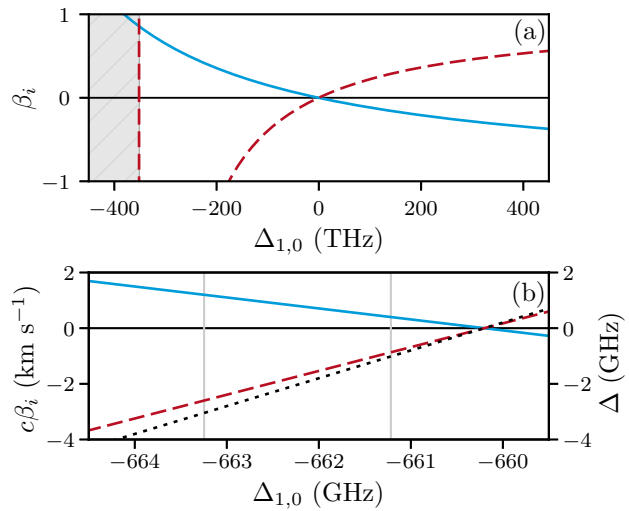}
	\vspace{-7mm}
	\caption{(a) Velocity distances  $\dist_1$ (solid) and 
	$\dist_2$ (dashed) between the Raman resonance 
	and the rogue resonances versus detuning $\Dizero$. The inadmissible range $\omega_2<0$, is shaded in gray.
	 (b)  Real velocities $c\dist_i$ and
	 the Doppler-shifted one-photon detuning $\Delta$ 
	 (dotted) on a small scale. 
	 The detunings for parameter set $\psa$ and  
	 $\psbb$  are marked with vertical lines (Table \ref{tab:laserparameters}). }
	\label{fig:resv}
\end{figure}

\vspace{-2mm}
The positions of the resonances are controlled by 
the laser frequencies.
Obviously, the Raman resonance at $\vres$ 
should be within the ion velocity distribution, also depicted \infig~\ref{fig:veldist}. In contrast, 
the rogue resonances at $v_i$ should be spread far apart.
Therefore, we want to determine laser frequencies, such that 
the resonance separations
\begin{equation}\label{eq:vresdiff}
\dist_{i} = \frac{v_i - \tvres}{c}
\end{equation}
are maximized. First, for a given Raman resonance velocity $\tvres$ \eqref{eq:vres}, one obtains a linear  frequency relation 
\begin{equation}
\omega_2(\omega_1;\tvres)=\frac{\omega_1\alpha_+-\omega_{mg}}{\alpha_-}, \quad \alpha_\pm =1\pm\frac{\tvres}{c},
\end{equation}
where we disregard ac-frequency shifts, deliberately.
Second, the distances between the resonances 
\begin{align}
\dist_1(\omega_1)&= \frac{\omega_{eg}}{\omega_1}-\alpha_+,\\
\dist_2(\omega_1) &=  \alpha_-\left(1-\frac{\omega_{em}}{\omega_1\alpha_+ - \omega_{mg}}\right),
\end{align}
are now functions of $\omega_1$, which is depicted in \fig\ref{fig:resv}. The requirement of positive laser frequencies
$\omega_2(\omega_1)>0$, leads to a lower limit for $\omega_1>  \omega_{mg}/\alpha_+$. 
For ultraviolet to near-infrared frequencies $|\dist_2|>|\dist_1|$. 
Therefore, we only need to maximize the distance 
$\dist_1$, within the range 
$-\alpha_+< \dist_1 < \alpha_+\omega_{em}/\omega_{mg}$. 
For detunings $|\Dizero| < \SI{1}{THz}$, 
the hyperbolic shape of the distance  
\begin{equation}
\dist_1(\Dizero)=-\frac{\tvres}{c}
-\frac{\Dizero}{\omega_{eg}}+
\frac{\Dizero^2}{\omega^2_{eg}}+\cdots
\end{equation}
is almost linear. Then, the maximal  distance of  $\dist_1$ 
is only limited by the available laser powers and interaction time. The time should last 
at least for one $\pi$-pulse $t_\pi$ \eqref{eq:tpi0} of a Raman transition, where maximal population transfer is achieved.
 This time will be derived in the next section and 
 is proportional to the Doppler shifted one-photon detuning 
 $\Delta$ and anti-proportional to the laser power. 
Therefore, $\Delta$ is also depicted in
\fig\ref{fig:resv} (b) and the values of $\Dizero$ for the parameter set $\psa$ and $\psbb$, Table\,\ref{tab:laserparameters} are highlighted.
 These parameter sets lead to two distinct velocity distances $c\dist_{1}^\psa =\SI{400}{\meter\per \second}$ and $c\dist_{1}^\psbb = \SI{1200}{\meter\per \second}$, keeping $\Delta$ small enough for the experimentally given interaction time and provided laser power.
The distances to resonance 2 are $c\dist_{2}^\psa = \SI{867}{\meter\per \second}$ and $c\dist_{2}^\psbb = \SI{2601}{\meter\per \second}$. 

\section{Time resolved population transfer}\label{sec:sol}\vspace{-2mm}
In principle, we want to optimize the  population transfer from the 
initial state $\ket{g}$ to the metastable state $\ket{m}$.
Therefore, the velocity averaged quantum expectation value of the observable $\hat{\sigma}_{mm}$,
\begin{align}\label{eq:mv}
\langle \popm(t,v)\rangle_v 
&=\int_0^\infty dv\, f(v)\,\text{Tr}[\hat{\sigma}_{mm}\hat{\rho}(t;v)]
\end{align}
should be maximized.
The uncertainty of the voltage measurement is defined by the logarithmic derivative \eqref{eq:U},
\begin{equation}\label{eq:sensitivity}
s=\frac{\Delta U}{U} =\frac{2\dpopm}{\vres}.
\end{equation}
 The smallest uncertainties are obtained for minimal velocity widths $\dpopm$ of the transferred metastable state's population.
Both objectives require the solution of the OBEs \eqref{blochL} 
\begin{align}
\label{eq:rhosol}
\rho(t;v)&= V(v) e^{\Lambda(v) t} V(v)^{-1} \rho(t=0;v),\\
\label{eq:evprob}
L(v) V(v)& = V(v) \Lambda(v),
\end{align}
for each velocity within the distribution $f(v)$.
Here $ \Lambda_i$ are the eigenvalues 
and $V(v)$ is the eigenmatrix of the Liouvillian matrix $L(v)$. 

We have implemented a numerical procedure to solve these equations for all velocities and obtain averages. We refer to this as the exact solution. 
However, in order to get insights in the underlying physical mechanisms, we will discuss in the following simple approximations that match the exact solution very well. 
These approximations emphasize  the relevance of the individual processes contributing cumulatively to the exact result. 
\subsection{Stimulated Raman transition}\vspace{-2mm}
\label{sec:raman}
For far-detuned lasers $\Delta \gg \Gamma_{i}$, the excited state stays nearly unpopulated, and spontaneous emission is not an issue. Hence, in a small regime around the resonant velocity $\vres$, the dynamics can be approximated by an effective two-level system, consisting of the ground and metastable states. This describes the process of the stimulated Raman transition.

We are dealing with laser linewidths $\gamma\sim \unit[10^2]{kHz}$, much larger Rabi frequencies $\Omega_i\sim\unit[10^2]{MHz}$ and even larger one-photon detunings $\Delta_{0,i}\sim\unit[10^2-10^3]{GHz}$. Around the Raman resonance, the two-photon detuning $\delta$ is very small, leading to the inequality 
\begin{equation}\label{eq:simplify}
\gamma,\delta\ll \Omega_i \ll\Delta_i.
\end{equation}
We will frequently make use of this relation.

\subsubsection{Coherent dynamics}\label{sec:coherent}
\begin{figure}[t]
	\centering
	\includegraphics{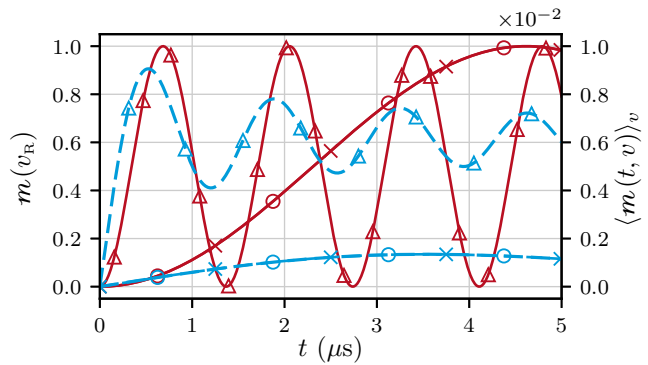}
	\vspace{-6mm}
	\caption{Rabi oscillations of the metastable state population $\popm(t,\vres)$ (red solid) for the resonant velocity, together with the velocity averaged population $\langle \popm(t,v)\rangle_v$ for $\dvel = \SI{50}{\meter\per\second}$ (blue dashed). Three parameter sets (Tab. \ref{tab:laserparameters}) are compared: $\psa$ ($\times$), $\psb$ ($\Circle$), and $\psc$ ( $\triangle$).}
	\label{fig:Rabi}
\end{figure}

\vspace{-2mm}
The ideal coherent dynamic is described by the Schrödinger equation
\begin{equation}
i \partial_t\ket{\psi} = 
(\mathcal{H}/\hbar-\varpi) \ket{\psi},
\end{equation}
with $\ket{\psi} = \psi_g \ket{g} + \psi_e\ket{e} + \psi_m\ket{m}$ and using the Hamilton matrix \eqref{eq:hamilton}. 
In order to apply the standard adiabatic elimination methods \cite{Brion2007} to eliminate the tiny excited-state population using $\dot{\psi}_e\ll\Delta \psi_e$, we have to transform to another frame.
This is accomplished by introducing the frequency shift $\varpi=\Delta +\delta/2$, leading only to an unobservable, global, dynamical phase. The resulting effective two-level system reads
\begin{equation}\label{eq:SG}
i\partial_t\!\begin{pmatrix} \psi_g \\ \psi_m \end{pmatrix}
=\begin{pmatrix}\!
\frac{\delta}{2}+\omega_{\text{ac}1}& \frac{\Or}{2}\\ 
\frac{\Or^\ast}{2}&  -\frac{\delta}{2}+\omega_{\text{ac}2}
\end{pmatrix}
\begin{pmatrix} \psi_g \\ \psi_m\end{pmatrix},
\end{equation}
with the Raman Rabi frequency and the ac-Stark shifts
\begin{equation}
\Or=\frac{ \Oi^\ast\Oii}{2\Delta} +\mathcal{O}\left(\tfrac{\delta}{ \Delta}\right), \label{eq:raman}
\qquad 
\omega_{\text{ac}i}=\frac{ |\Omega_1|^2}{4\Delta} +\mathcal{O}\left(\tfrac{\delta}{ \Delta}\right),\ 
\end{equation}
using the separation of the frequency scales \eqref{eq:simplify}. The two-level dynamics \eqref{eq:SG} can be solved by diagonalization as stated in Eq. (\ref{eq:rhosol},\,\ref{eq:evprob}). For the initial condition \mbox{$\psi_g(t=0)=1$, }
the metastable state's population reads
\begin{align}\label{eq:solSG}
\popmse(t,v)&=|\psi_m(t,v)|^2= 
\frac{|\Or|^2}{\Omega^2}\sin ^2\Big[\frac{\Omega t}{2}\Big], \\
\Omega&=\sqrt{|\Or|^2 + \tid^2}, \nonumber
\end{align}
with the effective detuning $\tid(v)= \underline{\Delta}_1-\underline{\Delta}_2$.
For weak saturation solution $\popmse(t,v)$ \eqref{eq:solSG} can be simplified with
\begin{align}\label{eq:Ramanapx}
\Or \approx\Ov\equiv \frac{\Oi^\ast\Oii}{2 \tiD}, 
\quad
\tid\approx\dur \equiv(k_1 + k_2) (v-\vres),
\end{align}
approximating the velocity dependent Rabi frequencies with their on-resonance values 
\eqref{eq:approx}.
\fig \ref{fig:Rabi} shows Rabi oscillations of the population of the metastable state for the resonant velocity $\popm(t,v=\vres)$  together with its velocity average $\langle \popm(t, v)\rangle_v$, calculated with \eqref{eq:mv}. In our Calcium experiment, ions emerge from the accelerator with a flat top velocity distribution $f(\dev) = 1/\dvel$ for $|\dev|\leq \dvel /2$, vanishing elsewhere, with the relative velocity $\dev = v-\vmean$.\\

Three different laser parameter sets, listed in Table \ref{tab:laserparameters}, are compared, demonstrating the essential impact of different laser frequencies and powers. Parameter sets $\psa$ and $\psb$ generate a $\pi$-pulse for the experimental transit time $\bar{\tau}=\unit[4.62]{\mu s}$. For the resonant velocity [$\dur(v=\vres)=0$] a complete population transfer is achieved. Applying \eqref{eq:Ramanapx} to \eqref{eq:solSG}, we obtain the $\pi$-pulse duration
\begin{equation}\label{eq:tpi0}
\tpizero\equiv t_\pi(\gamma=0)=\frac{\pi }{|\Ov|}.
\end{equation}
Parameter sets $\psa$ and $\psbb$ differ in the laser frequencies, resulting in vastly different distances between the  stimulated and the spontaneous Raman resonances \mbox{$\dist_1^\psb >\dist_1^\psa $,} as mentioned inSec.\,\ref{sec:resv}.  However, this does not affect the Rabi oscillations and $\langle \popm(t, v)\rangle_v$, because for the purely coherent population transfer via the Raman transition the spontaneous one is not an issue at all.
The parameter set $\psc$ provides the same laser frequencies as $\psb$, while the maximum laser power, available in the experiment, is applied. Therefore, $\langle \popm(t, v)\rangle_v$ is slightly enlarged, effectively due to power broadening. 
\begin{figure}[t]
	\centering
	\includegraphics{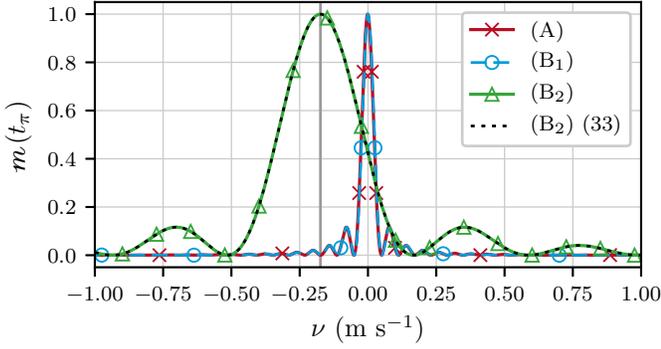}
	\vspace{-7mm}
	\caption{The velocity-dependent metastable-state population $\popm(t_\pi)$ after a $\pi$-pulse (\mbox{$t_\pi^\psa\! = t_\pi^\psb\! = \unit[4.62]{\mu s}$} and $t_\pi^\psc\! = \unit[0.68]{\mu s}$) is indiscernible for parameter sets $\psa$ and $\psb$. For maximal laser power $\psc$ the resonance is broadened as well as shifted. The approximation $\popmse$ \eqref{eq:sinc} matches the full solution.}
	\label{fig:veldisp_ana}
\end{figure}
This is apparent \infig \ref{fig:veldisp_ana}, depicting the velocity dispersion of the metastable state's population after a $\pi$-pulse.
 Using approximations \eqref{eq:Ramanapx}, and the expression for the $\pi$-pulse time \eqref{eq:tpi0} this population reads
\begin{align}\label{eq:sinc}
\popmse(\tpizero, v)=\frac{\pi^2}{4}\text{sinc}^2\!\Big[\frac{\pi \tilde{\Omega}}{2|\Ov|}\Big],
\end{align}
with $\text{sinc}[x]=\sin[x]/x$. It is plotted for $\psc$, matching the exact analytical solution \eqref{eq:solSG}.
The sinc$^2$-behavior is the typical response to constant interaction. For smooth temporal envelopes the side maxima vanish. Again the results for $\psa$ and $\psb$ show no difference. For $\psc$ the resonance is ac-Stark shifted ($\Omega_1^\psc \neq \Omega_2^\psc$) to $\dvres ^\psc=\vres-\vmean=\SI{-0.17}{\meter \per\second}$ as predicted by \eqref{eq:vres}. \\

We define the velocity width (FWHM) of $\popmse$ \eqref{eq:sinc} by the first zero of the sinc-function
\begin{equation}\label{eq:fwhmSG}
\dpopmse(\tpizero) = \sqrt{3} \frac{|\Ov|}{k_1+k_2}.
\end{equation}
For maximal laser power $\dpopm(t_\pi)$ increases from $\dpopm^\psb(t_\pi\!=\!\SI{4.62}{\micro \second}) =\SI{0.05}{\meter \per\second}[\SI{0.05}{\meter \per\second}]$ to $\dpopm^\psc(t_\pi\!=\!\SI{0.68}{\micro \second}) =\SI{0.32}{\meter \per\second}[\SI{0.34}{\meter \per\second}]$, where the results of approximation
 \eqref{eq:fwhmSG} are displayed in square brackets and provide very good predictions.
\subsubsection{Finite laser linewidths}\vspace{-2mm}
For finite laser linewidths, the ground-state decoherence must be considered. 
Adiabatic elimination of the fast coherences $\rho_{ge}$ and $\rho_{me}$ in the OBEs \eqref{eq:BEmasterpop} and \eqref{eq:BEmastercoh} leads to the following equations of motion in matrix representation $\dot{\rho} = L\rho$ with $\rho=(\rho_{gg},\rho_{gm},\rho_{mg},\rho_{mm})$ and
\begin{equation}\label{eq:M2ph}
L=\begin{pmatrix}
0 &i\frac{\Or^\ast}{2} & -i\frac{\Or}{2}&0 \\
i\frac{\Or}{2} &-\gamma-i\tid &0 &-i\frac{\Or}{2}\\
-i\frac{\Or^\ast}{2} &0 &-\gamma+i\tid&i\frac{\Or^\ast}{2}\\
0 &-i\frac{\Or^\ast}{2}&i\frac{\Or}{2} &0
\end{pmatrix}.
\end{equation}
Furthermore, we neglect the spontaneous emission rates as well as population of the excited state. In addition, we exploit the separation of the frequency scales \eqref{eq:simplify}.
 Applying the initial condition $\rho_{gg}(t=0)=1$, leads to the solution for the population of the metastable state
\begin{equation}\label{eq:sol2ph}
\popmph(t, v)=\frac{1}{2} + e^{-\vartheta t}(A \cos{\theta t}+ B\sin{\theta t}) - e^{2(\vartheta-\gamma)t}C.
\end{equation}
The stationary solution approaches $\lim_{t\rightarrow \infty}\popmph = 1/2$. The complex, velocity-dependent frequency and damping rate  $\theta$, $\vartheta$ as well as the coefficients $A$, $B$, and $C$ define damped oscillations (cf. Appendix\,\ref{app:coeff}).

\subsubsection{$\pi$-pulses for underdamped oscillations}\vspace{-2mm}
The solution \eqref{eq:sol2ph} includes regimes from underdamping to overdamping, depending on the ratio $\eta=\sqrt{3}|\Ov|/\gamma$. 
For $\eta>1$ and $v=\vres$, the coefficient $C(\vres)=0$. The oscillation frequency $\theta$ and damping rate 
$\vartheta$ are real.
In order to maximize the population transfer, we define 
 a $\pi$-pulse time $t_\pi$ for the \mbox{resonant velocity} \mbox{$v=\vres$} from the condition $\dot{\popm}_\text{R}(t_\pi,\vres)=0$ and obtain
\begin{align}\label{eq:tpi}
t_\pi&=\frac{1}{\theta}\cos^{-1}\left(\frac{A\theta + B\vartheta}{\sqrt{(A^2+B^2)(\theta^2+\vartheta^2)}}\right)\nonumber \\
&\approx \frac{\pi}{\theta}
= \frac{2 \sqrt{3} \pi  \xi}{\gamma\left(\xi^2-1 +\eta^2\right)},
\\
\xi&=\left[1-\frac{\eta^2}{2}\left(1- 
\sqrt{4\eta^2-3}\right)\right]^{1/3},\nonumber
\end{align}
using again the approximations $\Or\approx \tilde{\Omega}_\text{R}$ and $\tid\approx d$ \eqref{eq:Ramanapx}. 
In the limit $\gamma=0$, we recover  $\tpizero$ \eqref{eq:tpi0}.

For $t_\pi$, an upper bound for the velocity width of $\popmph(t_\pi,v)$ is given by the FWHM of 
\begin{equation}
e^{2(\vartheta-\gamma)t_\pi}C\approx \frac{1}{2}\frac{\dur^2}{\zeta^2+\dur^2}
\text{exp}\left[\frac{D\, t_\pi}{1+E\, \dur^2}\right],
\end{equation}
where $\zeta^2=\gamma^2(\eta^2-1)/3$ and $D$ and $E$ are velocity-independent functions, given in Appendix\,\ref{app:coeff}. We get
\begin{align}\label{eq:fwhm}
\dpopmph(t_\pi)&=\frac{2}{k_1+k_2}\sqrt{p+\sqrt{p^2+q}},\\
p&=\frac{\zeta^2E-Dt_\pi-\ln(2)}{2\ln(2)E},\ 
q= \frac{\zeta^2}{\ln(2)E}\nonumber,
\end{align}
using $\ln(d^2+\zeta^2)=\ln(d^2)+\zeta^2/d^2
+\mathcal{O}((\zeta^2/d^2)^2)$.
Together with the lower bound, provided by the limit of vanishing laser linewidths $\dpopmse$ \eqref{eq:fwhmSG}, the width is constrained by
\begin{equation}
\dpopmse (t_\pi)\leq\dpopm (t_\pi)\leq\dpopmph(t_\pi),
\end{equation}
for arbitrary $\eta>1$.
\subsubsection{Limit of large laser linewidths}\label{sec:rateequations}\vspace{-2mm}
In the limit $t\gg1/\gamma$ of \textit{large} laser linewidths and \textit{long} interaction times, the populations of the ground and metastable state can be approximated with the solutions of the rate equations derived from the two-level OBEs \eqref{eq:M2ph} with adiabatic elimination, using $\dot{\rho}_{gm} \ll \gamma  \rho_{gm}$,
\begin{equation}\label{eq:RE}
 \partial_t \begin{pmatrix}\rho_{gg}\\ \rho_{mm}\end{pmatrix}
= r
\begin{pmatrix}   -1 & 1 \\ 1 & -1 \end{pmatrix}
\begin{pmatrix}\rho_{gg}\\ \rho_{mm}\end{pmatrix}.
\end{equation}
The decay rate $r=\gamma|\Ov|^2/(\gamma^2 +\dur ^2)<r_0=r(d=0)$ 
involves the approximations $\Or\approx \tilde{\Omega}_\text{R}$ and $\tid\approx d$ \eqref{eq:Ramanapx} and the solution reads
\begin{align}\label{eq:solREa}
\popmre(t, v)&=\tfrac{1}{2}\left(1-e^{-r t}\right).
\end{align}
The velocity width of $\popmre$ follows with
\begin{align}\label{eq:fwhmRE}
\dpopmre(t) &=\frac{2\gamma}{k_1+k_2}\sqrt{ \frac{r_0 t}{\ln(1+\text{tanh}\frac{r_0 t}{2})}-1},
\end{align}
where $\dpopmre(t)\geq\gamma/(k_1+k_2)\ \forall\, r_0, t$. 
The presented rate equation limit is a good approximation, when the transient part \eqref{eq:sol2ph} vanishes, which is the case if  $\gamma t\gg1$, because $\gamma/2 \leq \vartheta <\gamma$.

\subsubsection{$\pi$-pulses for overdamped oscillations}\vspace{-2mm}
For $\eta<1$, the solution $\popmre$ \eqref{eq:solREa} is overdamped with $\popmre(t)<1/2\ \forall \ t$ . Therefore, we define the time of a $\pi$-pulse, when $\popmre(v=\vres)$ is saturated after several decay times \mbox{$t_\pi=n/r_0$} with $n>1$. We choose $n$ by the condition \mbox{$\lim_{\eta\rightarrow 1-}n/r_0=\lim_{\eta\rightarrow 1+}t_\pi$}
with $t_\pi(\eta>1)$ from Eq. \eqref{eq:tpi}, leading to
\begin{align}\label{eq:tpi2}
t_\pi=\frac{2\pi\gamma}{\sqrt{3}\spacefortildeB|\Ov|^2},\quad 0<\eta \leq 1.
\end{align}

\subsection{Spontaneous Raman transition}\label{sec:spont}\vspace{-2mm}
 The atomic transition between the ground and the excited state can be coupled resonantly, depending on the frequency of laser 1 and the ion velocity. In this limit, the population transferred into the metastable state can be approximated with the solution of the rate equations for the ground, excited, and metastable states, approximating the steady state of all coherences. Therewith, the OBEs \eqref{eq:BEmasterpop} and \eqref{eq:BEmastercoh} simplify to
 \begin{equation}\label{eq:M1ph}
 \partial_t\!
 \begin{pmatrix}
 \rho_{gg}\\ \rho_{mm}\\ \rho_{ee}
 \end{pmatrix}\negthickspace
 = \negthickspace
 \begin{pmatrix}
 -\tOi  & 0 & \Gamma_{eg} + \tOi\\
0  &  -\tOii  &  \Gamma_{em} + \tOii \\
 \tOi& \tOii & -\Gamma - \tOi -\tOii \\
 \end{pmatrix}\negthickspace
 \begin{pmatrix}
 \rho_{gg}\\ \rho_{mm}\\ \rho_{ee}
 \end{pmatrix}\!,
 \end{equation}
 with $R_i = \Gamma_i |\Omega_i|^2/(4 \Delta_i^2 + \Gamma_i^2)$. Additionally, we approximate $\rho_{gm}(t)\approx 0$,
 being important only in the regime of the transfer via the Raman transition. 
 For $t>1/\Gamma$ the population of the metastable state reads
\begin{align}\label{eq:sol1ph}
\popml(t, v)&=\frac{\frac{\Gamma_{em}}{\tOii}+1}{\frac{\Gamma_{em}}{\tOii}+\frac{\Gamma_{eg}}{\tOi}+3}\left(1-e^{-rt}\right),\\
r&=\frac{\tOi\Gamma_{em}+\tOii(\Gamma_{ge} +3\tOi)}{\Gamma+ 2 (\tOi+\tOii)}\nonumber.
\end{align}
 \subsection{Maximizing the population transfer}\vspace{-2mm}
 After studying the individual population transfer into the metastable state via the stimulated Raman transition $\popmph$ \eqref{eq:sol2ph} and the spontaneous population transfer $\popml$  \eqref{eq:sol1ph} separately, we will analyze the total population transfer now.
 We assume that we can incoherently combine both processes, as long as they do not interfere. The \textit{ad hoc} analytical ansatz 
 \begin{align}\label{eq:sol12}
\popmana(t, v) &= \tfrac{1}{w} \popmph(t, v) +\popml(t,v),\\
w &= \popmph(\tpizero, \vres) + \popml(\tpizero, \vres), \nonumber
 \end{align}
 superposes the populations such that $\popmana(\tpizero,\vres)=1$.
 \begin{figure}[t]
 	\centering
	\includegraphics{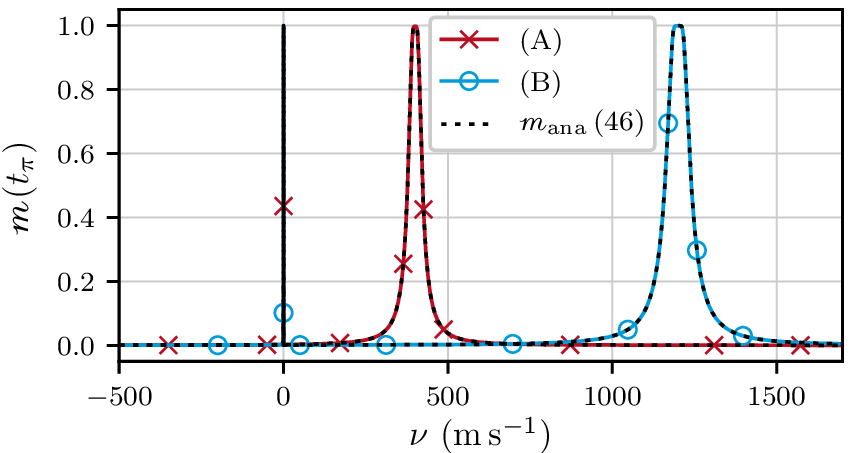}
 	\vspace{-6mm}
 	\caption{Velocity-dependent population of the metastable state after applying a $\pi$-pulse   $\popm(t_\pi)$, conforming with the analytical approximation \eqref{eq:sol12}. The narrow Raman resonance at $\dvres\approx\SI{0}{\meter\per\second}$, shown in details \infig\ref{fig:veldisp_ana} and  the broad resonance of laser 1 at $\dvi^\psa =\SI{400}{\meter\per\second}$ and  $\dvi^\psbb=\SI{1200}{\meter\per\second}$ are apparent.}
 	\label{fig:simb0v}
 \end{figure}

\subsubsection{Vanishing laser linewidths}
\begin{figure}[t]
	\includegraphics{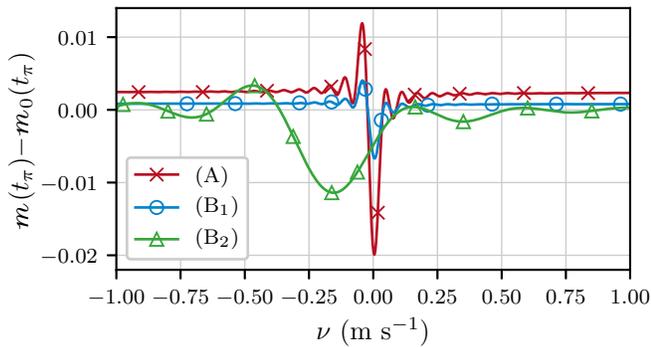}
	\vspace{-6mm}
	\caption{Velocity-dependent difference between the metastable state's population, considering ($\popm$) and neglecting ($\popmse$) spontaneous emission effects, after $t_\pi$.
	}\label{fig:simb0v_vres2}
\end{figure}

\vspace{-2mm}
Using the numerical solutions of the OBEs \eqref{eq:BEmasterpop} and \eqref{eq:BEmastercoh}, we can calculate the population distributions over a wide velocity range, depicted in \fig \ref{fig:simb0v}. The approximation $\popmana$ \eqref{eq:sol12} clearly matches the numerical results. The rogue resonance of spontaneous population transfer, located at $\dvi^\psa=\SI{400}{\meter \per\second}$ and $\dvi^\psbb =\SI{1200}{\meter \per\second}$, is clearly distinguishable from the narrow Raman resonance at $\dvres^\psa=\dvres^\psb\approx\dvres^\psc\approx\SI{0}{\meter \per\second}$. Obviously parameter sets $\psbb$ are more favorable, because here it can be ensured that for wider velocity distributions of the ions the transferred population into $\ket{m}$
 remains less influenced by the tail of the rogue resonance as for $\psa$.
However, for all parameter sets there is indeed a small deviation to the reference $\popmse$  \eqref{eq:solSG}, considering exclusively the coherent transfer via the Raman transition. This difference is depicted \infig \ref{fig:simb0v_vres2}. For $\psa$ the deviation and especially the roughly constant offset besides the Raman resonance  is clearly larger than for $\psb$ and $\psc$, because of a smaller separation of the rogue to the Raman resonance $\beta_1^\psa < \beta_1^\psbb$.  However, with enlarging the laser power $\psc$, especially the reduction exactly on the Raman resonance is as expected slightly enlarged again.
Nevertheless, the velocity width of the transferred population is quasi purely defined by the width of the Raman transition $\dpopmana = \dpopmph$, showing no differences from the results of Sec.~\ref{sec:coherent} [cf. $\dpopmse(\tpizero)$ \eqref{eq:fwhmSG} and \fig\ref{fig:veldisp_ana}].\\
\begin{figure}[t]
	\centering
	\includegraphics{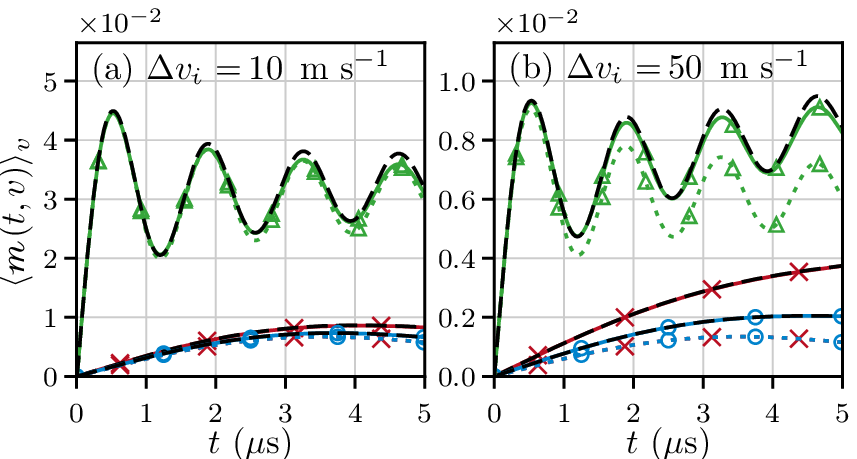}
	\vspace{-6mm}
	\caption{Time evolution of the velocity averaged metastable state's population  $\langle \popm(t,v)\rangle_v$ for two ion velocity widths (a) $\dvel=\SI{10}{\meter\per\second}$ and (b) $\dvel=\SI{50}{\meter\per\second}$ and the parameters $\psa$ (\textcolor{tud9c}{$\times$}), $\psb$ (\textcolor{tud2a}{$\Circle$}), and $\psc$ ($\textcolor{green}{\triangle}$). 
		The numerical solutions considering incoherent effects (solid), well predicted by the analytical approximation \eqref{eq:sol12} (black dashed), are compared to the analytic ones,  indicating the transferred population purely  due to the Raman transition \eqref{eq:solSG}  (dotted). }\label{fig:simb0mv}
\end{figure}
The negative impact of a \textit{small} distance $\beta_1$ can be further illustrated comparing the  velocity averaged population $\langle \popm(t,v)\rangle_v$  \eqref{eq:mv} \infig \ref{fig:simb0mv}. 
Already for the \mbox{as narrow} as possible initial ion velocity distribution with $\dvel = \SI{10}{\meter\per\second}$, the effect of the incoherent population transfer is discernible for $\psa$. For $\dvel= \SI{50}{\meter\per\second}$ and $\psa$ the total transferred population is more than twice of the population, transferred via the Raman transition.
So for $\psa$ this regime is already unsuitable to properly determine the ion velocity. 
However, the major portion of $\langle \popm(t,v) \rangle _v$ for $\psb, \psc$ results still from the narrow Raman transition. In addition, the necessary, larger Rabi frequencies are still reachable, viz., this represents a good working regime for available initial ion velocity widths.

For parameter sets $\psa$ and $\psb$ the analytical approximation $\popmana$ \eqref{eq:sol12} describes exactly the numerical results. Only for the maximum laser power within $\psc$ are there tiny deviations after the first $\pi$-pulse. 

\subsubsection{Finite laser linewidths}\vspace{-2mm}
Taking into account a non-vanishing laser linewidth ($\gamma=\Gamma_{gg}=\Gamma_{mm}=\SI{300}{\kilo\hertz}$ ) the velocity selectivity of the Raman transition is significantly affected. On the one hand, its width is enlarged, so the velocity determination is less exact. On the other hand, even on the resonance $\vres$ the maximum transfer efficiency of almost $100\%$ for $t_\pi$ cannot be reached any longer. Both effects are visible \infig \ref{fig:bvres} (a), depicting the velocity dependent population of the metastable state after time $\tpizero$. 
However, the Raman resonance is still clearly discernible and just as for $\gamma=0$ the analytic approximation $\popmana$ \eqref{eq:sol12} can predict the numerical, full solution. We have foregone plotting the results for parameters $\psa$, because they are very similar to $\psb$. However, the differences are not negligible, becoming apparent in \fig \ref{fig:spatio} (a), where the velocity average $\langle \popm(t,v)\rangle_v$ is visualized. Again the analytic approximation $\popmana$ \eqref{eq:sol12} gives reliable predictions; only for the maximum laser power and longer times are tiny deviations visible. 

\begin{figure}[t]
	\includegraphics[width=\columnwidth]{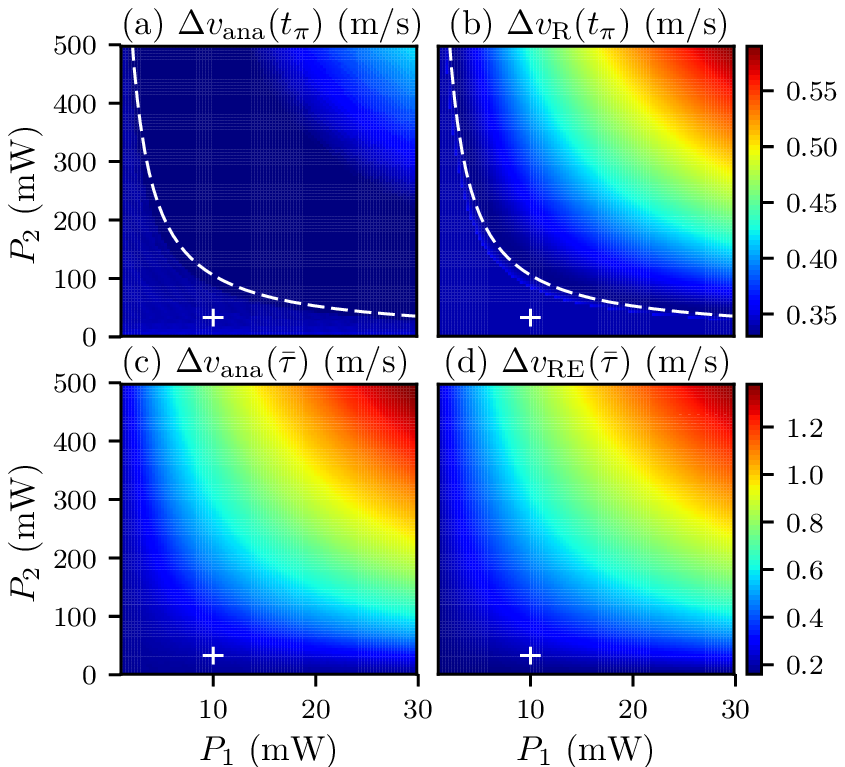}\\
	\caption{Velocity width $\dpopm$ (FWHM) of the metastable state's population depending on the laser powers $P_1,\ P_2$ for different times:
	after applying a $\pi$-pulse [(a),(b)] and $\bar{\tau}={\unit[4.62]{\mu s}}$ [(c),(d)]
	The FWHM of $\popmana$ \eqref{eq:sol12} [(a),(c)] is compared to the approximations $\dpopmph$ \eqref{eq:fwhm} (b) and $\dpopmre$ \eqref{eq:fwhmRE} (d). The crosses highlight the laser powers used with parameter set $\psb$ and the boarder between under- and overdamping is highlighted (dashed white).}
\label{fig:fwhm}
\end{figure}

The disappearance of Rabi oscillations indicates that the laser linewidths and the interaction time are sufficient to yield the rate equation limit. Only for $\psc$ is the time of a $\pi$-pulse discernible with $t_\pi^{\psc} = \SI{0.70}{\micro \second}$, slightly different from $\tpizero^\psc=\SI{0.68}{\micro \second}$.  The width of the velocity dispersion is $\dpopm(t_\pi) = \SI{0.42}{\meter \per\second}$, where the analytical approximations $\dpopmse(\tpizero)$ \eqref{eq:fwhmSG} and $\dpopmph(t_\pi)$ \eqref{eq:fwhm} provide an appropriate range 
$\dpopm(t_\pi) \in [\dpopmse(\tpizero),\dpopmph(t_\pi) ]  = [0.34, 0.60 ]\SI{}{\meter \per\second}$.
The $\pi$-pulse times for the parameter sets $t_\pi^{\psa} \approx t_\pi^{\psb}  =\SI{14.8}{\micro \second}$ are much larger than $\tpizero^\psa =\tpizero^\psb= \SI{4.62}{\micro \second}$, demonstrating the overdamping for $\psa$, $\psb$ with $\gamma=\SI{300}{\kilo\hertz}$. Hence, we get for the interaction time $\bar{\tau}=\SI{4.62}{\micro \second}$, $\dpopm(\bar{\tau}) = [ 0.22, 0.22, 1.34]\si{\meter\per\second}$ for $[\psa, \psb, \psc]$, demonstrating the broadening of the Raman transition due to finite laser linewidths and well predicted by the analytical approximation $\dpopmre(\bar{\tau}) = [0.21, 0.21, 1.38]\si{\meter\per\second}$ \eqref{eq:fwhmRE}.
In this way, in particular for $\psc$, the total amount of the metastable-state population is 
substantially enlarged. At the same time, the ratio of population transferred into the metastable state via the Raman transition to the whole population 
\begin{equation}\label{eq:ratio}
\mu =\frac{\langle \popmph(t,v,r)\rangle_{v, r}}{\langle \popmana(t,v,r)\rangle_{v, r}}, 
\end{equation}
depicted in Figs.\,\ref{fig:spatio}(e) and \ref{fig:spatio}(g), remains larger in comparison to the simulations with $\gamma=0$. 
The comparison of the parameter sets demonstrates the compelling necessity of a careful choice of laser frequencies, providing a sufficiently large distance between Raman and rogue resonance $\beta_1$, especially for wider velocity distributions of the ions. It is worth mentioning, that the ratio $\mu$ decreases with time, more crucial for the idealized scenario with $\gamma=0$ than for $\gamma\neq0$. Therefore, it is important to carefully choose the interaction time in combination with the laser powers, achieving a significant absolute population amount and  simultaneously keeping a reliable ratio, optimally $\mu\rightarrow1$ but at least $\mu>0.5$. To identify the maximum laser powers to reach a prescribed maximum uncertainty of the velocity determination, the widths $\dpopm(t_\pi)$ as well as $\dpopm(\bar{\tau})$ are depicted \infig\ref{fig:fwhm}.
Indeed, $\dpopmana(t_\pi)$ \eqref{eq:fwhm} provides an upper bound for $\dpopm (t_\pi)$, and  due to $\gamma\bar{\tau}=9\gg1$, $\dpopmre(\bar{\tau})$ \eqref{eq:fwhmRE} matches the actual results $\dpopmana (\bar{\tau})$ very well. It is worth mentioning that the relative deviation between $\dpopmana$ and $\dpopm$ of the numerical solution of the OBEs \eqref{eq:BEmasterpop} and  \eqref{eq:BEmastercoh}, not depicted here, is in the lower single-digit percentage range with no qualitative difference.

\subsubsection{Spatial intensity variations}\label{sec:spatio}\vspace{-2mm}
\begin{figure}[t]
	\includegraphics[width=\columnwidth]{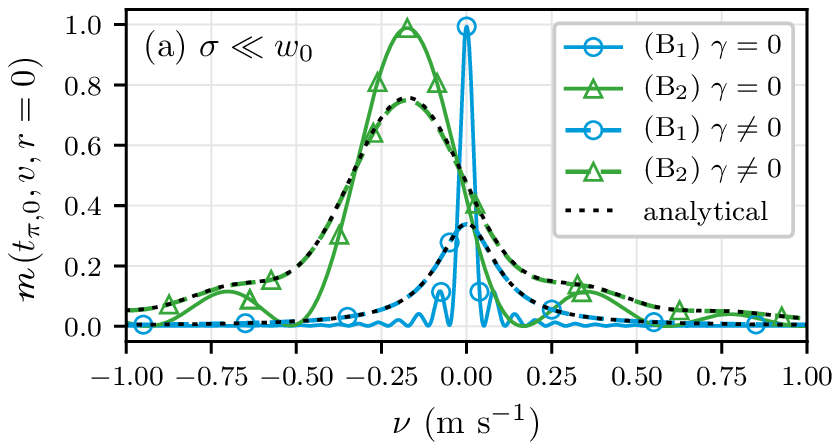}\\
	\vspace{-10mm}
	\includegraphics[width=\columnwidth]{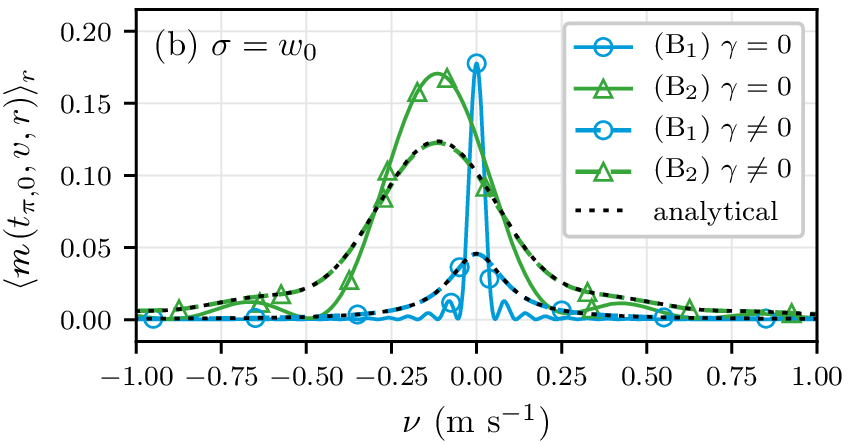}
	\vspace{-6mm}
	\caption{Velocity-dependent population of the metastable state after $\tpizero$ for different spatial distributions: (a) no intensity variations according to $\sigma \ll w_0$, and (b) averaged over the spatial intensity variations for an ion beam widths $\sigma = w_0$. Considering finite laser linewidths $\gamma=\unit[300]{kHz}$ (dashed) is compared to $\gamma=0$ (solid). The analytic approximation \eqref{eq:sol12} (dotted) matches the simulation results.}
	\label{fig:bvres}
\end{figure}
\begin{figure*}
\includegraphics{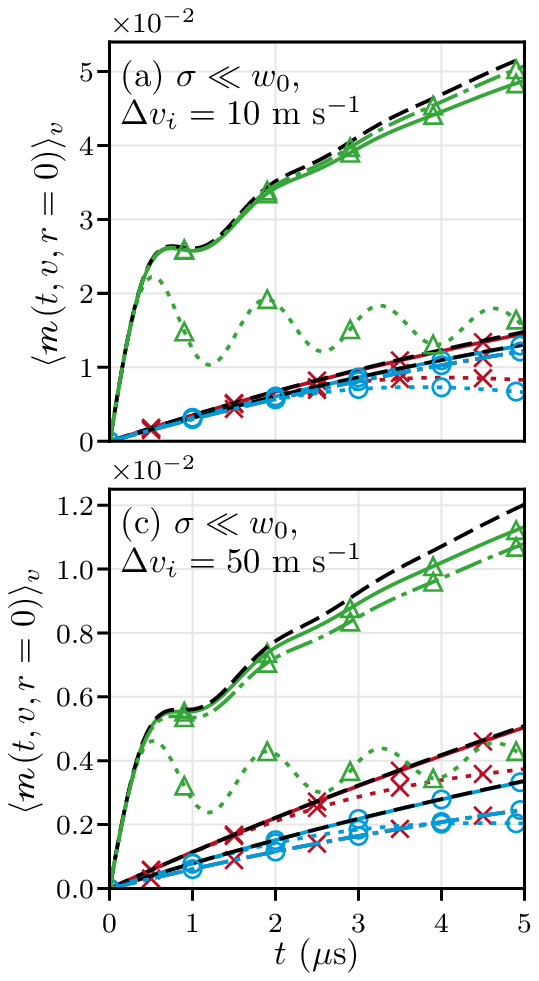}\hspace{-2mm}
\includegraphics{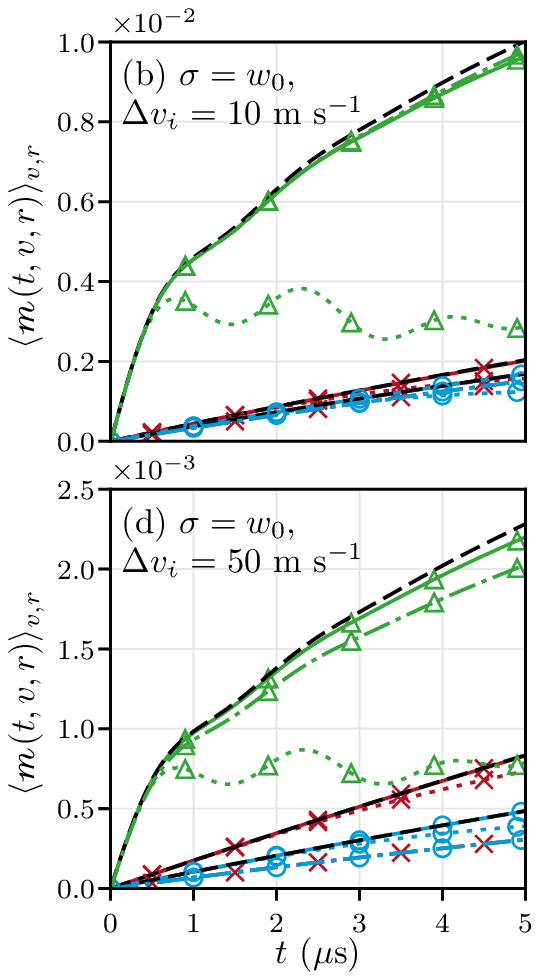}\hfill
\includegraphics{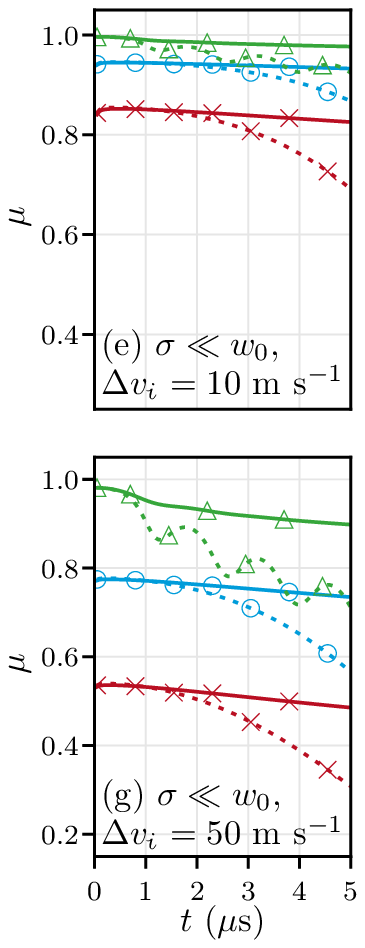}
\includegraphics{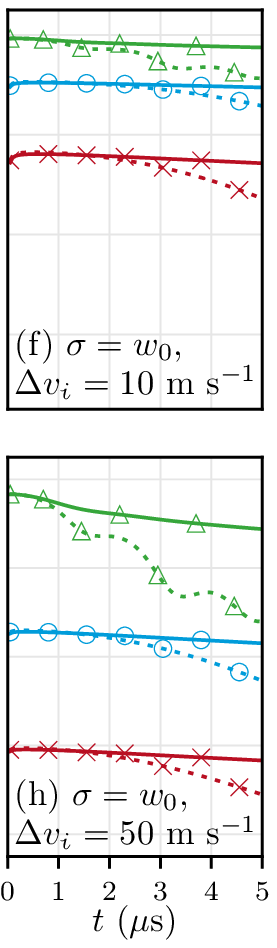}
	\vspace{-4mm}
	\caption{
		(a) - (d): Time evolution of the metastable state's population summed over a certain velocity width.
		(e) - (h): Time-dependent ratio $\mu$ \eqref{eq:ratio} of the population transferred into the metastable state via the Raman transition to the whole transferred population, including spontaneous population transfer.
		The upper row considers an initial ion velocity width $\dvel=\SI{10}{\meter\per\second}$ and the lower row $\dvel=\SI{50}{\meter\per\second}$. 
		Two spatial intensity distributions are analyzed: neglecting spatial variations according to $\sigma \ll w_0$ [(a),(c),(e),(g)] and averaging over the spatial distributions for an ion beam width $\sigma = w_0$ [(b),(d),(f),(h)] for the parameter sets: $\psa$ (\textcolor{tud9c}{$\times$}), $\psb$ (\textcolor{tud2a}{$\Circle$}) and $\psc$ ($\textcolor{green}{\triangle}$), latter scaled with a factor of $0.5$ in (a) - (d). 
		(a) - (d): The numerical results for $\Gamma\neq 0, \gamma=\unit[300]{kHz} $ (solid) are well predicted by the analytic approximation for the full solution $\eqref{eq:sol12}$ (dashed black). The analytic approximation $\eqref{eq:sol2ph}$ (dash-dotted) gives the populations purely transferred via the Raman transition.
		The numerical solutions for $\gamma=0$ (dotted) are given for the sake of completeness.
	   (e) - (h): Considering a finite laser linewidth $\gamma=\unit[300]{kHz}$ (solid) is compared to neglecting it $\gamma=0$ (dotted).}
	\label{fig:spatio}
\end{figure*}
So far we approximate the lasers as plane waves with no spatial dependences. In reality, they are collimated Gaussian laser beams with beam waist $w_0$ and
\begin{equation}
\Omega(r) =\Omega_0\,  e^{-r^2/w_0^2}.
\end{equation}
The ions are also spatially inhomogeneously distributed and assumed to be Gaussian with width $\sigma$,
\begin{equation}
g(r) = \frac{1}{2\pi\sigma^2}e^{-r^2/2 \sigma^2}.
\end{equation}
We average over the cross-sectional area of the ion and the laser beams, assuming that the respective maxima of all are perfectly overlapped
\begin{equation}
\begin{split}
\langle \popm(t, v, r)\rangle _{ r} &= 2\pi \int_{0}^{\infty} dr\,  r\, g(r)\,  \popm(t, v, r).
\end{split}
\label{eq:poweraverage}
\end{equation}
We solve this integral numerically with Gauss-Laguerre quadrature \cite{Abramowitz1964}.
Finally, the population of $\ket{m}$ is additionally averaged over the velocity distribution according to \eqref{eq:mv}.\\

We compare two scenarios, the results of which are depicted in Figs.\,\ref{fig:bvres} and \ref{fig:spatio}.
For	$\sigma\ll w_0$, the calcium beam is much narrower than the laser beam, corresponding to the case we assumed so far in which all ions experience the same Rabi frequency [Figs. \ref{fig:bvres}(a), \ref{fig:spatio}(a), \ref{fig:spatio}(c), \ref{fig:spatio}(e), and \ref{fig:spatio}(g)]. For $\sigma=w_0$, the calcium beam is broader than the laser beam, viz. some ions are not affected at all, representing the current experimental scenario [Figs. \ref{fig:bvres}(b), \ref{fig:spatio}(b), \ref{fig:spatio}(d), \ref{fig:spatio}(f), and \ref{fig:spatio}(h)].\\
For $\sigma = w_0$, essentially, $\langle \popm(\tpizero, v, r)\rangle_{r}$ (\fig \ref{fig:bvres}) and $\langle \popm(t, v, r)\rangle_{v, r}$ (\fig \ref{fig:spatio}) are reduced overall.

Further away from the center $r=0$, the Rabi frequencies are reduced and therewith the Raman resonance velocity $\vres$ is shifted. This effect is more crucial, the larger $\Omega_2$ is. Averaging over the results for different $r$,  for $\psc$ the velocity of maximum transfer efficiency is shifted from $\dvres=\vres-\vmean=\SI{-0.17}{\meter\per\second}$ to $\dvres=\SI{-0.11}{\meter\per\second}$. However, the  resonance of the Raman transition is still visible. In addition, the considered spatial intensity variations lead to a small reduction of the ratio $\mu$ \eqref{eq:ratio}, primarily for $\psa$ and $\psb$.

\section{Conclusion}\vspace{-2mm}
Our calculations show, that ion velocity classes with widths as low as $\Delta v=\SI{0.2}{\meter\per\second}$ can be transferred into the metastable state via the Raman transition, achieving a significant population proportion $\langle \popm(t, v, r)\rangle_{v, r} = 10^{-3}-10^{-2}$.
Thereby it is important to carefully choose the laser frequency combination to ensure that the transferred population into the metastable state originates mainly by the Raman process and not by incoherent spontaneous emission processes, when laser 1 couples resonantly to $\ket{e}$. This also supports an initially narrow ion velocity distribution.

The idealized case of infinitely sharp laser linewidths ($\gamma=0$) and an ion beam much smaller than the laser beams defines the smallest reachable FWHMs of the ion velocity distribution in the metastable state
with $\dpopm(\tpizero)=[0.05,0.05, 0.32]\,\si{\meter\per\second}$, for parameter sets $[\psa, \psb,\psc]$. With \eqref{eq:sensitivity} this results in a voltage-width  \mbox{$s=[0.4,0.4,  2.4]\, \textrm{ppm}$} on the sub-ppm level for $\psa$ and $\psb$.
The analytical expressions for the resonance velocity of the Raman transition $\vres$ \eqref{eq:vres}, the population of the metastable state after applying a $\pi$-pulse $\popmse(\tpizero, v)$ \eqref{eq:sinc}, and the corresponding FWHM $\dpopmse(\tpizero) $ \eqref{eq:fwhmSG} give reliable predictions.

Considering finite laser linewidths, the analytical approximation for the velocity- and time-dependent metastable-state population $\popmana$ \eqref{eq:sol12} matches the full numerical solution very well. In addition, 
the approximations for the velocity width after applying a $\pi$-pulse $\dpopmph(t_\pi)$ \eqref{eq:fwhm} and in the rate equation limit $\dpopmre(t_\pi)$ \eqref{eq:fwhmRE} are also suitable. Therewith, almost exact results for large parameter regimes are predicted with simultaneously small computational effort. Moreover, the presented analytical models lead to physical insights, verified by the numerical results.

Finite laser linewidths lead to a significant broadening of the Raman transition. With $\gamma=\SI{300}{\kilo\hertz}$ the velocity width of $\popm$ is enlarged to $\dpopm(\tpizero)= [0.22,0.22, 0.43]\,\si{\meter\per\second}$, leading to $s=[1.7, 1.7,3.3]\, \textrm{ppm}$, still on the ppm level. Note that the width of the distribution does not represent the ultimate limit of the determination of the resonance velocity $\vres$. Therefore, sub-ppm high-voltage measurements are still attainable.
The related width in the frequency domain is $\Delta f= (k_1+k_2) \dpopm /(2\pi)=[0.8,0.6, 1.6]\,\si{\mega\hertz}$, much smaller than the natural linewidth $\Gi=\SI{23.396}{\mega\hertz}$.
The Raman transition therefore has the potential to provide a significant reduction in uncertainty. Moreover, it avoids additional uncertainties caused by varying and unknown momentum transfers in multiple resonant excitations along the $4s\rightarrow 4p$ transition and the subsequent spontaneous decay in the current measurement scheme. The momentum transfer during the Raman transition is very small
and exactly defined as $h(k_1+k_2)^2/2m=\SI{69}{\kilo\hertz}$ in direction of laser 2. Therefore, it can be taken into account in the analysis process. 

In addition, with the velocity acceptance of the Raman transition, we can approximate the maximum angle between ion and laser beams, where ions can be just transferred into the metastable state \mbox{$\alpha \leq \arccos[(\vres -\dpopm/2)/(\vres+\dvel/2)]\approx\SI{6}{\milli\radian}$} for all parameter sets.
In comparison, the natural linewidth induces a much larger angle \mbox{$\alpha' \leq \arccos[(v_1-\Gamma_{eg}/(2 k_1))/(v_1+\dvel/2)]=\SI{11}{\milli\radian}$.}

Spatial intensity variations of both the laser beams and the ion beam mainly reduce slightly (less than an order of magnitude) the transfer efficiency for all velocities. Therewith, the velocity width of the transferred population is approximately not affected. Note, that for maximum laser powers the velocity of maximum population transfer is slightly shifted in comparison to infinitely large laser beams. 
However, considering these experimental imperfections, the resonance of the Raman transition is still clearly identifiable. This demonstrates the feasibility of high-voltage measurements using coherent Raman spectroscopy on the ppm level, under realistic conditions.
Finally, high-precision collinear laser spectroscopy from metastable states might also profit from such a Raman velocity filter due to the reduction in Doppler linewidth.

\section{Acknowledgements}\vspace{-2mm}
We thank Kristian König and Jörg Krämer for fruitful discussions and helpful suggestions.
A.N. and R.W. acknowledge support from the DLR German Aerospace Center with funds provided by the Federal Ministry for Economic Affairs and Energy (BMWi) under Grant No. 50WM1957. 
W.N. acknowledges support from the Helmholtz International Center for 
FAIR within the LOEWE program by the State of Hesse.
\appendix

\section{Raman transition in incoherent beams}\vspace{-2mm}
\label{raman}
We consider a Raman transition 
with counter-propagating lasers
$\vec{E}_i(t,\rv)=\Re{\vec{\epsilon}_i\mathcal{E}_i e^{i (\vec{k}_i \rv-\omega_i t)}}$ as shown \infig\ref{fig:laserion}.
The coordinate $\rv$ refers to the laboratory frame $S$. 
Assuming that the ion beam and the lasers are aligned
along the z-direction, we can specify the ion velocity 
$\vec{v} = v\vec{e}_z$ and the laser wave vectors 
$\vec{k}_1 = -k_1 \vec{e}_z$ and $\vec{k}_2 = k_2 \vec{e}_z$. 
During the interaction of the laser pulses, no relevant transversal motion occurs, which is why we will discuss only the one-dimensional evolution in the z-direction.

In the optical domain, the electric dipole interaction in the rotating-wave approximation is dominant \cite{Cohen2004}.
Thus, the 
Hamilton operator of an ion with quantized  canonical coordinates
$[\zop,\pzop]=i\hbar$ is
\begin{align}\label{eq:H0}
H(t)=& \frac{\pzop^2}{2m}+ \hbar(\omega_g \siggg +\omega_m\sigmm +\omega_e\sigee)\\ 
 & +\hbar\left( \sigeg \tfrac{\Omega_1}{2} e^{i\phi_1}+ 
 \sigem\tfrac{\Omega_2}{2} e^{i \phi_2}+  \textrm{H.c.}\right),\nonumber
\end{align}
with phases $\phi_1=-k_1 \zop -\omega_1 t$, $\phi_2=k_2 \zop - \omega_2 t$.
It accounts for the kinetic and the internal energy of the ion, where $\hbar \omega_i$ is the energy of state 
$\ket{i}$. The electronic transition operators are
$\hat\sigma_{ij} = \dyad{i}{j}$.
The complete internal and motion state $\hat{\varrho}(t)$ of the ionic beam evolves according to 
\begin{align}
\dot{\hat{\varrho}}=-i[H/\hbar,\hat{\varrho}].
\end{align}
One eliminates the ballistic evolution 
by the transformation 
$\hat{\varrho}=\exp(-it \pzop^2/2m\hbar)
\hat{\varrho'}
\exp(it \pzop^2/2m\hbar)$. Then, the Liouville-von-Neumann equation
$\dot{\varrho}'=-i[H'/\hbar,\hat{\varrho}']$ reads
\begin{align} 
H'(t)/\hbar=&
\omega_g \siggg +\omega_m\sigmm +\omega_e\sigee\\ 
  &+\left( 
  	\sigeg \tfrac{\Omega_1}{2} e^{i \phi'_1}+ 
 	 \sigem\tfrac{\Omega_2}{2} e^{i \phi'_2}+  
 	 \textrm{H.c.}\right),\nonumber
\end{align} 
with Doppler shifted phases 
$\phi'_1=-k_1 \zop -(\omega_1+k_1 \pzop/m)t$
and $\phi'_2=k_2 \zop - (\omega_2-k_2 \pzop/m)t$.
The motional state of the ionic beam smoothly extends over a large phase-space area $\Delta \zop\Delta \pzop\gg \hbar/2$. The photon recoils $\hbar k_i \ll \Delta \pzop$ are tiny compared to the momentum width and the recoil energy $\hbar^2 (k_1+k_2)^2/2m\ll \hbar \Gamma_i , \hbar \Omega_i$ is tiny compared to the level shifts or widths, however the Doppler shifts $k_i \Delta \pzop\gg \Gamma_i,\Omega_i$ are significant.  Hence, we use the classical approximation 
$(\zop,\pzop)\rightarrow (z,p_z=m v)$ 
of kinetic theory \cite{Dalibard1985}.
Consequently, the full quantum state 
$\hat{\varrho}'(t)\rightarrow \hat{\rho}'(t;z,v)$ is replaced by an internal state operator denoting the motional variables to the role of parameters. 
Using the Doppler shifted frequencies $\omega'_i$ from 
Eq.~\eqref{eq:omdop}, the phases read
$\phi'_1=-k_1 z -\omega'_1t$ and $\phi'_2=k_2 z - \omega'_2t$ and
the Hamilton operator in the classical approximation $H'\rightarrow\mathcal{H}'$ is given by
\begin{align} 
\mathcal{H}'(t)/\hbar=&\omega_g \siggg +\omega_m\sigmm +\omega_e\sigee\\ 
  &+\left( \sigeg \tfrac{\Omega_1}{2} 
  e^{i \phi'_1}+ 
  \sigem\tfrac{\Omega_2}{2} e^{i\phi'_2}
  +  \textrm{H.c.}\right).\nonumber
\end{align} 
The remaining optical and spatial oscillations are eliminated  by the transformation $\hat{\rho}'=U(t,z)\hat{\rho} U^\dag(t,z)$, with
\begin{align}
U(t,z) =&e^{-i \omega_e t -i\phi'_1 \hat\sigma_{gg}
+i \phi'_2  \hat\sigma_{mm}}.
\end{align}
This results in the Liouville von Neumann equation for the semi classical state $\dot{\hat{\rho}}=-i[\mathcal{H}/\hbar,\hat{\rho}]$,
with
\begin{align} 
\frac{\mathcal{H}}{\hbar}=&
\Delta_1\siggg +\Delta_2\sigmm
+\left( 
  	\sigeg \tfrac{\Omega_1}{2} +\sigem\tfrac{\Omega_2}{2}+ 
  	 \text{H.c.}
  	\right),
\end{align} 
with a one-photon detuning $\Delta_1=\omega'_1-\omega_{eg}$ for laser 1 
and $\Delta_2=\omega'_{2}-\omega_{em}$ for laser 2.
With this approximation for the Hamiltonian evolution, we have essentially disregarded any photon recoil effects.  One can apply the same arguments to the spontaneous contributions of the Lindblad equation \cite{Dalibard1985}.

\section{Bloch matrix}\vspace{-2mm}
\label{app:blochm}
Representing the master equation \eqref{eq:master} in 
 the sorted basis 
$\{\ket{g},\ket{e},\ket{m}\}$ and arranging the matrix elements as 
linear arrays $\rho=(\rho_g,\rho_e,\rho_m)$ with
$\rho_i=(\rho_{ig}, \rho_{ie},\rho_{im})$, one finds the Bloch equation 
\eqref{blochL} with a
Bloch matrix 
\begin{widetext}
\begin{align}
L=&
\begin{pmatrix}
L_{gg}&L_{ge}&0\\
L_{eg} & L_{ee} & L_{em}\\
0 & L_{me} & L_{mm}
\end{pmatrix}
=i \left(
\begin{array}{ccc ccc ccc}
 0 & \frac{\Omega _1}{2} & 0 & -\frac{\Omega_1^\ast}{2} & -i
   \Gamma _{\text{eg}} & 0 & 0 & 0 & 0 \\
 \frac{\Omega_1^\ast}{2} & i \Gamma _1-\Di  &
   \frac{\Omega_2^\ast}{2} & 0 & -\frac{\Omega_1^\ast}{2} & 0 & 0 & 0 & 0 \\
 0 & \frac{\Omega_2}{2} & i \gamma-\delta   & 0 & 0 & 
 -\frac{\Omega_1^\ast}{2} & 0 & 0 & 0 \\
 -\frac{\Omega_1}{2} & 0 & 0 & \Di+i \Gamma _1  & 
 \frac{\Omega_1}{2} & 0 & -\frac{\Omega_2}{2} & 0 & 0 \\
 0 & -\frac{\Omega_1}{2} & 0 & \frac{\Omega_1^\ast}{2} & i \Gamma
    & \frac{\Omega_2^\ast}{2} & 0 & -\frac{\Omega_2}{2} & 0 \\
 0 & 0 & -\frac{\Omega_1}{2} & 0 & \frac{\Omega_2}{2} & \Dii +i \Gamma _2 &
   0 & 0 & -\frac{\Omega_2}{2} \\
 0 & 0 & 0 & -\frac{\Omega_2^\ast}{2} & 0 & 0 & \delta +i \gamma 
   & \frac{\Omega_1}{2} & 0 \\
 0 & 0 & 0 & 0 & -\frac{\Omega_2^\ast}{2} & 0 & \frac{\Omega_1^\ast}{2} & i \Gamma _2-\Delta_2  & 
 \frac{\Omega_2^\ast}{2} \\
 0 & 0 & 0 & 0 & -i \Gamma _{\text{em}} & -\frac{\Omega_2^\ast}{2}
   & 0 & \frac{\Omega_2}{2} & 0 \\
\end{array}
\right),
\end{align}
\end{widetext}
with $\delta=\Delta_1-\Delta_2$.
It exhibits the block structure of two coupled two-level systems.
The Bloch equations define an initial value problem with $\rho(t=0)=(\rho^0_g,\rho^0_e,\rho^0_m)=(1,0,0,0,0,0,0,0,0)$. The Laplace transform
\begin{align}
\varrho(s)=\int_0^\infty dt \,e^{-st} \rho(t)
\end{align}
is ideally suited to transform the system of differential equations with initial values to an algebraic equation
\begin{align}
\begin{pmatrix}
{\mathcal{G}_{g}^{0}}^{-1}&-L_{ge}&0\\
0 & {\mathcal{G}^{0}_{e}}^{-1} & -L_{em}\\
0 & 0 & {\mathcal{G}^0_{m}}^{-1}
\end{pmatrix}
\begin{pmatrix}
 \varrho_g\\
 \varrho_e\\
 \varrho_m
\end{pmatrix}=
\begin{pmatrix}
\rho_{g}^0\\
L_{eg}\varrho_g\\
L_{me}\varrho_e
\end{pmatrix},
\end{align}
where $\mathcal{G}^0_{\lambda}(s)=(s-L_{\lambda\lambda})^{-1}$ is 
the resolvent matrix. The formal inversion of the Bloch matrix is
facilitated by the block structure and by backward substitution. This leads
to the explicit solution
\begin{align}\label{eq:laplace}
\varrho_g(s)&=\mathcal{G}_{g}(s)\rho_{g}^0, &
\mathcal{G}_{g}^{-1}&={\mathcal{G}^0_{g}}^{-1}-L_{ge}\mathcal{G}_{e}L_{eg},\notag\\
\varrho_e(s)&=\mathcal{G}_{e}(s)L_{eg}\varrho_{g}(s), & 
\mathcal{G}_{e}^{-1}&={\mathcal{G}^0_{e}}^{-1}-L_{em}\mathcal{G}^0_{m}L_{me},\notag\\
\varrho_m(s)&=\mathcal{G}^0_{m}(s)L_{me}\rho_{e}(s), & \varrho_i&=(\rho_{ig}, \rho_{ie},\rho_{im}).
\end{align}
One can find the stationary solution   using the final value theorem of the Laplace transformation 
\begin{equation}\label{eq:laplacestatsol}
\rho_{mm}^\infty\equiv\lim_{t\rightarrow\infty}\rho_{mm}(t)=\lim_{s\rightarrow 0}s \varrho_{mm}(s).
\end{equation}

  The Laplace transform can also be used to approximate the initial growth rate $\dot{\rho}_{mm}(t=0)$ of the population of the metastable state. This provides insights into the contributions of different processes of population transfer. Therefore, \infig \ref{fig:poptransschema} the processes generating population in $\ket {m}$ are schematically visualized. 
Following this scheme and starting initially with the whole population in the ground state, the Laplace transform, considering only the initial processes in perturbation theory, denoted by $\tilde{\varrho}_{mm}$, results in 
 \begin{align}
 \label{eq:apprate}
  \begin{split}
 \tilde{\varrho}_{mm}&=G_{mm}\left[
 \left(\Gamma_{em}+\frac{|\Omega_2|^2}{2}\Re{G_{me}}\right)\tilde{\varrho}_{ee}\right.\\
 &\phantom{=G_{mm}}\left.+\frac{|\Omega_2|^2|\Omega_1|^2}{8}\Re{G_{me}G_{mg}G_{eg}}G_{gg}
 \right],
 \end{split}\\
 \tilde{\varrho}_{ee}&= \frac{|\Omega_1|^2}{2}G_{ee}\Re{G_{eg}}G_{gg},\nonumber
  \end{align}
 \begin{align*}
 G_{gg}^{-1}&=s,  &  G_{ee}^{-1}&=s+\Gamma , & G_{eg}^{-1}&=s-i(\Di)+\Gamma_1,\notag\\
G_{mm}^{-1}&=s,  &  G_{gm}^{-1}&=s+i\delta+\gamma,&  G_{em}^{-1}&=s-i\Dii+\Gamma_2.
 \end{align*}
The initial growth rate of the metastable-state population is then given by $r_m=\lim_{s\rightarrow 0}s^2\tilde{\varrho}_{mm}(s)$.

\section{Parameters}\vspace{-2mm}
\label{app:parameters}
Relevant spectroscopic data for  \ca are given in Table~\ref{tab:ionparameters}, while  
laser parameters are specified in Table~\ref{tab:laserparameters}.
\begin{table*}[t] 
	\caption{
	Parameters for \ca\ transitions between the states $\ket{g}$, $\ket{e}$, $\ket{m}$, of the configurations 
	$4s\,^2S_{1/2}$, 
	$4p\,^2P_{3/2}$, $3d\,^2D_{5/2}$.}
	\label{tab:ionparameters}
	\begin{tabular}{ l l cc  rr r}
		Quantity && Symbol && Value && Reference\\
		\hline
		\hline
		Mass && $m$ &&  \SI{39.962042286(22)}{u} &&\cite{Wang2017, NIST18}\\
		\hline
		Angular transition frequency && $\omge$  &&  
		$2\pi\times\SI{761.905 012 599 (82)}{THz}$&& \cite{Shi2017}\\
		Natural linewidth (FWHM)  & &$\Gamma\uge$  & &  
		$2\pi\times\SI{23.396}{MHz}$& &\cite{NIST18}\\
		Lifetime  && $\tau\uge=\Gamma^{-1}\uge$  && \SI{6.8}{ns}&&\\
		Transition dipole matrix element && $\langle J\!=\!\tfrac{1}{2}||er||J'\!=\!\tfrac{3}{2}\rangle$  &&\SI{2.301129e-29}{\coulomb \meter}&&\\
		\hline
		Angular transition frequency  && $\omme$ && $2\pi\times 
		\SI{350.862 882 823 (82)}{THz}$&& \cite{Guan2015,Shi2017}\\
		Natural linewidth (FWHM)  && $\Gamma\ume$ & & $2\pi\times\SI{1.576}{MHz}$ &&\cite{NIST18}\\
		Lifetime  & & $\tau\ume=\Gamma\ume^{-1}$  && \SI{101}{ns} &&\\
		Transition dipole matrix element & &
		$\langle J\!=\!\tfrac{5}{2}||er||J'\!=\!\tfrac{3}{2}\rangle$ && \SI{1.250998e-29}{\coulomb \meter}&&\\
		\hline
		Acceleration voltage  && $U$ &&\SI{14}{\kilo\volt}&&\\
		Mean velocity && $\vmean$ && \SI{260}{\kilo\meter\per\second}&&\\
		Initial width of velocity distribution  (FWHM) & &$\dvel$ && 
		\SIrange{10}{100}{\meter\per \second}&&\\
		\hline
		\hline
	\end{tabular}
\end{table*}

\begin{table*}[t] 
	\caption{Laser parameters, where the Rabi frequency is calculated with \eqref{eq:rabi}, \eqref{eq:deff}.}
	\label{tab:laserparameters}
	\begin{tabular}{l l c c  r r r r r}
		Quantity & & Symbol && Parameter set A& &Parameter set B$_1$ && Parameter set B$_2$ \\
		\hline\hline
		Laser 1 & & & && &&&\\
		\hline
		\textbf{Frequency} && $f_1$  && $761.24\mathbf{3\, 795\, 50}\,\si{\tera\hertz}$ && 
		\multicolumn{3}{c}{
			\textcolor{lightgray}{\rule[1mm]{6mm}{0.4pt}}\hspace{3mm}
			$761.24\mathbf{1\, 765\, 92}\,\si{\tera\hertz}$
			\hspace{3mm}\textcolor{lightgray}{\rule[1mm]{6mm}{0.4pt}}}\\
		Wavelength && $\lambda_1$ &&\SI{393.8192}{\nano\meter} &&
		\multicolumn{3}{c}{
			\textcolor{lightgray}{\rule[1mm]{6mm}{0.4pt}}\hspace{10.8mm}\hspace{3mm}
			\SI{393.8203}{\nano\meter}
			\hspace{3mm}\textcolor{lightgray}{\rule[1mm]{6mm}{0.4pt}}}\\
		Wave number && $k_1$ && \SI{15.954490}{\per\micro\meter}&&  
		\multicolumn{3}{c}{
			\textcolor{lightgray}{\rule[1mm]{6mm}{0.4pt}}\hspace{5.2mm}\hspace{3mm}
			\SI{15.954448}{\per\micro\meter}
		\hspace{3mm}\textcolor{lightgray}{\rule[1mm]{6mm}{0.4pt}}}\\
		\textbf{Power} && $P_1$ & &\textbf{\SI{3.29}{\milli\watt}} &  &
		\textbf{\SI{10}{\milli\watt}} && \textbf{\SI{30}{\milli\watt}} \\
		Rabi frequency && $\Oi$ && $2\pi\times\SI{14.828}{MHz}$ &&
		$2\pi\times \SI{25.852}{MHz}$& &	$2\pi\times \SI{44.777}{MHz}$\\
		Beam radius && $w_0$  && \multicolumn{5}{c}{\textcolor{lightgray}{\rule[1mm]{2.5cm}{0.4pt}}\hspace{5.mm}\SI{1.7}{\milli\meter}\hspace{5.mm}\textcolor{lightgray}{\rule[1mm]{2.5cm}{0.4pt}}}\\
		Linewidth && $\Gamma_{gg}$ && \multicolumn{5}{c}{\textcolor{lightgray}{\rule[1mm]{2.5cm}{0.4pt}}\hspace{4mm} \SI{300}{kHz}\hspace{4mm}\textcolor{lightgray}{\rule[1mm]{2.5cm}{0.4pt}}}\\
		\hline
		\hline
		Laser 2&&&&&&&& \\
		\hline
		\textbf{Frequency} && $f_2$ && $351.16\mathbf{6\, 422\, 00}\,\si{\tera\hertz}$ & &
		\multicolumn{3}{c}{
			\textcolor{lightgray}{\rule[1mm]{6mm}{0.4pt}}\hspace{3.2mm}
			$351.16\mathbf{4\, 388\, 90}\si{\tera\hertz}$
		\hspace{3.2mm}\textcolor{lightgray}{\rule[1mm]{6mm}{0.4pt}}}\\
		Wavelength && $\lambda_2$ && \SI{853.7048}{\nano\meter} &&
	    \multicolumn{3}{c}{ \textcolor{lightgray}{\rule[1mm]{6mm}{0.4pt}}\hspace{3.2mm}
	    	\hspace{10.2mm}\SI{853.7097}{\nano\meter}
    		\hspace{3.2mm}\textcolor{lightgray}{\rule[1mm]{6mm}{0.4pt}}}\\
		Wave number && $k_2$ &&
		\SI{7.359904}{\per\micro\meter} &&
		 \multicolumn{3}{c}{ 	\textcolor{lightgray}{\rule[1mm]{6mm}{0.4pt}}\hspace{3.2mm}
		 	\hspace{6.2mm}\SI{7.359861}{\per\micro\meter}
	 		\hspace{3.2mm}\textcolor{lightgray}{\rule[1mm]{6mm}{0.4pt}}}\\
		\textbf{Power} && $P_2$ & &\textbf{\SI{11.13}{\milli\watt}}& &
		\textbf{\SI{33}{\milli\watt}}&&  \textbf{\SI{500}{\milli\watt}}\\
		Rabi frequency && $\Oii$ && $2\pi\times \SI{14.827}{MHz}$ 
		&& $2\pi\times\SI{25.531}{MHz}$ && $2\pi\times\SI{99.379}{MHz}$\\
		Beam radius && $w_0$  &&
		\multicolumn{5}{c}{\textcolor{lightgray}{\rule[1mm]{2.5cm}{0.4pt}}\hspace{5.5mm}\SI{1.7}{\milli\meter}\hspace{5.5mm}\textcolor{lightgray}{\rule[1mm]{2.5cm}{0.4pt}}}\\
		Linewidth && $\Gamma_{mm}$ & &
		\multicolumn{5}{c}{\textcolor{lightgray}{\rule[1mm]{2.5cm}{0.4pt}}\hspace{4mm} \SI{300}{kHz}\hspace{4mm} \textcolor{lightgray}{\rule[1mm]{2.5cm}{0.4pt}}}\\
		\hline
		\hline
	\end{tabular}
\end{table*}

The listed Rabi frequency, defining the interaction strength, can be calculated with the total laser power $P$ and the effective dipole moment $d_\text{eff}$  to \cite{ShoreChap}
\begin{equation}\label{eq:rabi}
\Omega_0=\frac{|d_\text{eff}|}{\hbar} \sqrt{\frac{4 P }{\pi \epsilon_0 c w_0 ^2}},
\end{equation}
with the laser beam waist $w_0 = \text{FWHM}/\sqrt{2\ln 2}$, the vacuum permittivity $\epsilon_0$, and the speed of light $c$.
Due to the non-existent nuclear spin of the considered level configurations, there is no hyperfine splitting and consequently the lasers interact with the $J\rightarrow J'$ transition.
In addition, the lasers are linearly polarized, wherefore they interact only with one (of three) component of the dipole operator and the effective coupling strength is \cite{ShoreChap}
\begin{equation}\label{eq:deff}
|d_\text{eff}|^2 = \frac{1}{3}  |\langle J || e\vec{r} ||J'\rangle|^2.
\end{equation}
The numerical  values of the reduced matrix elements, listed in Table\,\ref{tab:ionparameters}, can be calculated from the lifetime \cite{ShoreChap}
\begin{equation}
\frac{1}{\tau}=\frac{\omega_0^3}{3\pi\epsilon_0 \hbar c^3}\frac{2J+1}{2J'+1}|\langle J||e\vec{r}||J'\rangle|^2.
\end{equation}
\section{Coefficients of $\popmph$ \eqref{eq:sol2ph} and $\dpopmph$ \eqref{eq:fwhm}}\label{app:coeff}\vspace{-2mm}
The analytic approximation for metastable-state population, transferred via the Raman transition $\popmph$, results in \eqref{eq:sol2ph},
with velocity-dependent coefficients
\begin{gather}
\begin{alignat*}{3}
\vartheta &= \frac{1}{12}\left(8\gamma +x + \frac{y}{x}\right), 
&\ \  & \ \  & \theta &=  \frac{x^2 -y}{4\sqrt{3}x},\\
A&=\frac{1}{3}\frac{m n+ o p}{m^2+o^2}, 
& & &B&=\frac{1}{3}\frac{mp -no}{m^2+o^2},
\end{alignat*}\\
\begin{alignat*}{3}
C&=\frac{36 x^2\tid^2 /\gamma^2+[x(x+2)+y]^2}{6(x^4+x^2y+y^2)}, 
& & &m&=\sqrt{3}(x^4-4y^2),\\
n&= \sqrt{3}(y-x^2)[x(x-4)+y],
& & &o&=(x^2-y)^2,
\end{alignat*}\\
\begin{align*}
	p&=4xy(x-1)-y^2+x^2[8(1+9\tid^2/\gamma^2) - x(4+x)],\\
	x&= \left( z +\sqrt{z^2-y^3}\right)^{1/3},\\
	y&=4\big(1 -3\tfrac{\tid^2+\Or^2}{\gamma^2}\big),\qquad\quad
	 z=8\big(1+9\tfrac{\tid^2-\Or^2/2}{\gamma^2}\big).
\end{align*}
\end{gather}

The corresponding width $\dpopmph$ can be analytically approximated for $t_\pi$ resulting in \eqref{eq:fwhm}
with coefficients 
\begin{gather}
\begin{align*}
D&= 3 F(d=0),\qquad
E=
\frac{1}{2} \frac{d^2}{d d^2} 
\frac{ 9F(d=0)}{3F(d)-2}\bigg|
_{\Or=\Ov, d=\Gamma},\\
F&=\frac{x^2+y}{6 x}\big|_{\tid=d}.
\end{align*}
\end{gather}

\bibliographystyle{apsrev4-2}
\bibliography{library_paper15_neu}

\begin{thebibliography}{38}%
\makeatletter
\providecommand \@ifxundefined [1]{%
 \@ifx{#1\undefined}
}%
\providecommand \@ifnum [1]{%
 \ifnum #1\expandafter \@firstoftwo
 \else \expandafter \@secondoftwo
 \fi
}%
\providecommand \@ifx [1]{%
 \ifx #1\expandafter \@firstoftwo
 \else \expandafter \@secondoftwo
 \fi
}%
\providecommand \natexlab [1]{#1}%
\providecommand \enquote  [1]{``#1''}%
\providecommand \bibnamefont  [1]{#1}%
\providecommand \bibfnamefont [1]{#1}%
\providecommand \citenamefont [1]{#1}%
\providecommand \href@noop [0]{\@secondoftwo}%
\providecommand \href [0]{\begingroup \@sanitize@url \@href}%
\providecommand \@href[1]{\@@startlink{#1}\@@href}%
\providecommand \@@href[1]{\endgroup#1\@@endlink}%
\providecommand \@sanitize@url [0]{\catcode `\\12\catcode `\$12\catcode
  `\&12\catcode `\#12\catcode `\^12\catcode `\_12\catcode `\%12\relax}%
\providecommand \@@startlink[1]{}%
\providecommand \@@endlink[0]{}%
\providecommand \url  [0]{\begingroup\@sanitize@url \@url }%
\providecommand \@url [1]{\endgroup\@href {#1}{\urlprefix }}%
\providecommand \urlprefix  [0]{URL }%
\providecommand \Eprint [0]{\href }%
\providecommand \doibase [0]{https://doi.org/}%
\providecommand \selectlanguage [0]{\@gobble}%
\providecommand \bibinfo  [0]{\@secondoftwo}%
\providecommand \bibfield  [0]{\@secondoftwo}%
\providecommand \translation [1]{[#1]}%
\providecommand \BibitemOpen [0]{}%
\providecommand \bibitemStop [0]{}%
\providecommand \bibitemNoStop [0]{.\EOS\space}%
\providecommand \EOS [0]{\spacefactor3000\relax}%
\providecommand \BibitemShut  [1]{\csname bibitem#1\endcsname}%
\let\auto@bib@innerbib\@empty
\bibitem [{\citenamefont {Smekal}(1923)}]{Smekal1923}%
  \BibitemOpen
  \bibfield  {author} {\bibinfo {author} {\bibfnamefont {A.}~\bibnamefont
  {Smekal}},\ }\href {https://doi.org/10.1007/BF01576902} {\bibfield  {journal}
  {\bibinfo  {journal} {Naturwissenschaften}\ }\textbf {\bibinfo {volume}
  {11}},\ \bibinfo {pages} {873} (\bibinfo {year} {1923})}\BibitemShut
  {NoStop}%
\bibitem [{\citenamefont {Raman}\ and\ \citenamefont
  {Krishnan}(1928)}]{Raman1928}%
  \BibitemOpen
  \bibfield  {author} {\bibinfo {author} {\bibfnamefont {C.~V.}\ \bibnamefont
  {Raman}}\ and\ \bibinfo {author} {\bibfnamefont {K.~S.}\ \bibnamefont
  {Krishnan}},\ }\href {https://doi.org/10.1038/121501c0} {\bibfield  {journal}
  {\bibinfo  {journal} {Nature}\ }\textbf {\bibinfo {volume} {121}},\ \bibinfo
  {pages} {501} (\bibinfo {year} {1928})}\BibitemShut {NoStop}%
\bibitem [{\citenamefont {Raman}(1965)}]{Raman1965}%
  \BibitemOpen
  \bibfield  {author} {\bibinfo {author} {\bibfnamefont {C.}~\bibnamefont
  {Raman}},\ }in\ \href {https://doi.org/10.1016/B978-1-4831-9745-6.50001-5}
  {\emph {\bibinfo {booktitle} {Nobel Lectures, Physics 1922-1941}}}\ (\bibinfo
   {publisher} {Elsevier},\ \bibinfo {year} {1965})\BibitemShut {NoStop}%
\bibitem [{\citenamefont {Ferraro}\ \emph {et~al.}(2003)\citenamefont
  {Ferraro}, \citenamefont {Nakamoto}, \citenamefont {Brown}, \citenamefont
  {Ferraro}, \citenamefont {Nakamoto},\ and\ \citenamefont
  {Brown}}]{Ferraro2003}%
  \BibitemOpen
  \bibfield  {author} {\bibinfo {author} {\bibfnamefont {J.~R.}\ \bibnamefont
  {Ferraro}}, \bibinfo {author} {\bibfnamefont {K.}~\bibnamefont {Nakamoto}},
  \bibinfo {author} {\bibfnamefont {C.~W.}\ \bibnamefont {Brown}}, \bibinfo
  {author} {\bibfnamefont {J.~R.}\ \bibnamefont {Ferraro}}, \bibinfo {author}
  {\bibfnamefont {K.}~\bibnamefont {Nakamoto}},\ and\ \bibinfo {author}
  {\bibfnamefont {C.~W.}\ \bibnamefont {Brown}},\ }\href
  {https://doi.org/10.1016/B978-012254105-6/50001-9} {\emph {\bibinfo {title}
  {{Introductory Raman Spectroscopy}}}}\ (\bibinfo  {publisher} {Academic
  Press},\ \bibinfo {year} {2003})\BibitemShut {NoStop}%
\bibitem [{\citenamefont {Aspect}\ \emph {et~al.}(1988)\citenamefont {Aspect},
  \citenamefont {Arimondo}, \citenamefont {Kaiser}, \citenamefont
  {Vansteenkiste},\ and\ \citenamefont {Cohen-Tannoudji}}]{arimondo88}%
  \BibitemOpen
  \bibfield  {author} {\bibinfo {author} {\bibfnamefont {A.}~\bibnamefont
  {Aspect}}, \bibinfo {author} {\bibfnamefont {E.}~\bibnamefont {Arimondo}},
  \bibinfo {author} {\bibfnamefont {R.}~\bibnamefont {Kaiser}}, \bibinfo
  {author} {\bibfnamefont {N.}~\bibnamefont {Vansteenkiste}},\ and\ \bibinfo
  {author} {\bibfnamefont {C.}~\bibnamefont {Cohen-Tannoudji}},\ }\href
  {https://doi.org/10.1103/PhysRevLett.61.826} {\bibfield  {journal} {\bibinfo
  {journal} {Phys. Rev. Lett.}\ }\textbf {\bibinfo {volume} {61}},\ \bibinfo
  {pages} {826} (\bibinfo {year} {1988})}\BibitemShut {NoStop}%
\bibitem [{\citenamefont {Kaufman}(1976)}]{Kaufman1976}%
  \BibitemOpen
  \bibfield  {author} {\bibinfo {author} {\bibfnamefont {S.~L.}\ \bibnamefont
  {Kaufman}},\ }\href {https://doi.org/10.1016/0030-4018(76)90267-4} {\bibfield
   {journal} {\bibinfo  {journal} {Opt. Commun.}\ }\textbf {\bibinfo {volume}
  {17}},\ \bibinfo {pages} {309} (\bibinfo {year} {1976})}\BibitemShut
  {NoStop}%
\bibitem [{\citenamefont {Wing}\ \emph {et~al.}(1976)\citenamefont {Wing},
  \citenamefont {Ruff}, \citenamefont {Lamb},\ and\ \citenamefont
  {Spezeski}}]{Wing1976}%
  \BibitemOpen
  \bibfield  {author} {\bibinfo {author} {\bibfnamefont {W.~H.}\ \bibnamefont
  {Wing}}, \bibinfo {author} {\bibfnamefont {G.~A.}\ \bibnamefont {Ruff}},
  \bibinfo {author} {\bibfnamefont {W.~E.}\ \bibnamefont {Lamb}},\ and\
  \bibinfo {author} {\bibfnamefont {J.~J.}\ \bibnamefont {Spezeski}},\ }\href
  {https://doi.org/10.1103/PhysRevLett.36.1488} {\bibfield  {journal} {\bibinfo
   {journal} {Phys. Rev. Lett.}\ }\textbf {\bibinfo {volume} {36}},\ \bibinfo
  {pages} {1488} (\bibinfo {year} {1976})}\BibitemShut {NoStop}%
\bibitem [{\citenamefont {Schinzler}\ \emph {et~al.}(1978)\citenamefont
  {Schinzler}, \citenamefont {Klempt}, \citenamefont {Kaufman}, \citenamefont
  {Lochmann}, \citenamefont {Moruzzi}, \citenamefont {Neugart}, \citenamefont
  {Otten}, \citenamefont {Bonn}, \citenamefont {{Von Reisky}}, \citenamefont
  {Spath}, \citenamefont {Steinacher},\ and\ \citenamefont
  {Weskott}}]{Schinzler1978}%
  \BibitemOpen
  \bibfield  {author} {\bibinfo {author} {\bibfnamefont {B.}~\bibnamefont
  {Schinzler}}, \bibinfo {author} {\bibfnamefont {W.}~\bibnamefont {Klempt}},
  \bibinfo {author} {\bibfnamefont {S.~L.}\ \bibnamefont {Kaufman}}, \bibinfo
  {author} {\bibfnamefont {H.}~\bibnamefont {Lochmann}}, \bibinfo {author}
  {\bibfnamefont {G.}~\bibnamefont {Moruzzi}}, \bibinfo {author} {\bibfnamefont
  {R.}~\bibnamefont {Neugart}}, \bibinfo {author} {\bibfnamefont {E.-W.}\
  \bibnamefont {Otten}}, \bibinfo {author} {\bibfnamefont {J.}~\bibnamefont
  {Bonn}}, \bibinfo {author} {\bibfnamefont {L.}~\bibnamefont {{Von Reisky}}},
  \bibinfo {author} {\bibfnamefont {K.~P.~C.}\ \bibnamefont {Spath}}, \bibinfo
  {author} {\bibfnamefont {J.}~\bibnamefont {Steinacher}},\ and\ \bibinfo
  {author} {\bibfnamefont {D.}~\bibnamefont {Weskott}},\ }\href
  {https://doi.org/10.1016/0370-2693(78)90224-1} {\bibfield  {journal}
  {\bibinfo  {journal} {Phys. Lett. B}\ }\textbf {\bibinfo {volume} {79}},\
  \bibinfo {pages} {209} (\bibinfo {year} {1978})}\BibitemShut {NoStop}%
\bibitem [{\citenamefont {Neugart}(1985)}]{Neugart1985}%
  \BibitemOpen
  \bibfield  {author} {\bibinfo {author} {\bibfnamefont {R.}~\bibnamefont
  {Neugart}},\ }\href {https://doi.org/10.1007/BF02354810} {\bibfield
  {journal} {\bibinfo  {journal} {Hyperfine Interact.}\ }\textbf {\bibinfo
  {volume} {24}},\ \bibinfo {pages} {159} (\bibinfo {year} {1985})}\BibitemShut
  {NoStop}%
\bibitem [{\citenamefont {Otten}(1989)}]{Otten1989}%
  \BibitemOpen
  \bibfield  {author} {\bibinfo {author} {\bibfnamefont {E.~W.}\ \bibnamefont
  {Otten}},\ }in\ \href
  {https://doi.org/https://doi.org/10.1007/978-1-4613-0713-6_7} {\emph
  {\bibinfo {booktitle} {Treatise Heavy-Ion Sci.}}},\ Vol.~\bibinfo {volume}
  {8},\ \bibinfo {editor} {edited by\ \bibinfo {editor} {\bibfnamefont {D.~A.}\
  \bibnamefont {Bromley}}}\ (\bibinfo  {publisher} {Plenum Press},\ \bibinfo
  {address} {New York},\ \bibinfo {year} {1989})\ Chap.~\bibinfo {chapter} {7},
  p.\ \bibinfo {pages} {517}\BibitemShut {NoStop}%
\bibitem [{\citenamefont {Blaum}\ \emph {et~al.}(2013)\citenamefont {Blaum},
  \citenamefont {Dilling},\ and\ \citenamefont
  {N{\"{o}}rtersh{\"{a}}user}}]{Blaum2013}%
  \BibitemOpen
  \bibfield  {author} {\bibinfo {author} {\bibfnamefont {K.}~\bibnamefont
  {Blaum}}, \bibinfo {author} {\bibfnamefont {J.}~\bibnamefont {Dilling}},\
  and\ \bibinfo {author} {\bibfnamefont {W.}~\bibnamefont
  {N{\"{o}}rtersh{\"{a}}user}},\ }\href
  {https://doi.org/10.1088/0031-8949/2013/T152/014017} {\bibfield  {journal}
  {\bibinfo  {journal} {Phys. Scr.}\ }\textbf {\bibinfo {volume} {T152}},\
  \bibinfo {pages} {014017} (\bibinfo {year} {2013})}\BibitemShut {NoStop}%
\bibitem [{\citenamefont {Campbell}\ \emph {et~al.}(2016)\citenamefont
  {Campbell}, \citenamefont {Moore},\ and\ \citenamefont
  {Pearson}}]{Campbell2016}%
  \BibitemOpen
  \bibfield  {author} {\bibinfo {author} {\bibfnamefont {P.}~\bibnamefont
  {Campbell}}, \bibinfo {author} {\bibfnamefont {I.~D.}\ \bibnamefont
  {Moore}},\ and\ \bibinfo {author} {\bibfnamefont {M.~R.}\ \bibnamefont
  {Pearson}},\ }\href {https://doi.org/10.1016/j.ppnp.2015.09.003} {\bibfield
  {journal} {\bibinfo  {journal} {Prog. Part. Nucl. Phys.}\ }\textbf {\bibinfo
  {volume} {86}},\ \bibinfo {pages} {127} (\bibinfo {year} {2016})}\BibitemShut
  {NoStop}%
\bibitem [{\citenamefont {Neugart}\ \emph {et~al.}(2017)\citenamefont
  {Neugart}, \citenamefont {Billowes}, \citenamefont {Bissell}, \citenamefont
  {Blaum}, \citenamefont {Cheal}, \citenamefont {Flanagan}, \citenamefont
  {Neyens}, \citenamefont {N{\"{o}}rtersh{\"{a}}user},\ and\ \citenamefont
  {Yordanov}}]{Neugart2017}%
  \BibitemOpen
  \bibfield  {author} {\bibinfo {author} {\bibfnamefont {R.}~\bibnamefont
  {Neugart}}, \bibinfo {author} {\bibfnamefont {J.}~\bibnamefont {Billowes}},
  \bibinfo {author} {\bibfnamefont {M.~L.}\ \bibnamefont {Bissell}}, \bibinfo
  {author} {\bibfnamefont {K.}~\bibnamefont {Blaum}}, \bibinfo {author}
  {\bibfnamefont {B.}~\bibnamefont {Cheal}}, \bibinfo {author} {\bibfnamefont
  {K.~T.}\ \bibnamefont {Flanagan}}, \bibinfo {author} {\bibfnamefont
  {G.}~\bibnamefont {Neyens}}, \bibinfo {author} {\bibfnamefont
  {W.}~\bibnamefont {N{\"{o}}rtersh{\"{a}}user}},\ and\ \bibinfo {author}
  {\bibfnamefont {D.~T.}\ \bibnamefont {Yordanov}},\ }\href
  {https://doi.org/10.1088/1361-6471/aa6642} {\bibfield  {journal} {\bibinfo
  {journal} {J. Phys. G Nucl. Part. Phys.}\ }\textbf {\bibinfo {volume} {44}},\
  \bibinfo {pages} {064002} (\bibinfo {year} {2017})}\BibitemShut {NoStop}%
\bibitem [{\citenamefont {Wendt}\ \emph {et~al.}(1999)\citenamefont {Wendt},
  \citenamefont {Blaum}, \citenamefont {Bushaw}, \citenamefont {Gr{\"{u}}ning},
  \citenamefont {Horn}, \citenamefont {Huber}, \citenamefont {Kratz},
  \citenamefont {Kunz}, \citenamefont {M{\"{u}}ller}, \citenamefont
  {N{\"{o}}rtersh{\"{a}}user}, \citenamefont {Nunnemann}, \citenamefont
  {Passler}, \citenamefont {Schmitt}, \citenamefont {Trautmann},\ and\
  \citenamefont {Waldek}}]{Wendt1999}%
  \BibitemOpen
  \bibfield  {author} {\bibinfo {author} {\bibfnamefont {K.}~\bibnamefont
  {Wendt}}, \bibinfo {author} {\bibfnamefont {K.}~\bibnamefont {Blaum}},
  \bibinfo {author} {\bibfnamefont {B.~A.}\ \bibnamefont {Bushaw}}, \bibinfo
  {author} {\bibfnamefont {C.}~\bibnamefont {Gr{\"{u}}ning}}, \bibinfo {author}
  {\bibfnamefont {R.}~\bibnamefont {Horn}}, \bibinfo {author} {\bibfnamefont
  {G.}~\bibnamefont {Huber}}, \bibinfo {author} {\bibfnamefont {J.~V.}\
  \bibnamefont {Kratz}}, \bibinfo {author} {\bibfnamefont {P.}~\bibnamefont
  {Kunz}}, \bibinfo {author} {\bibfnamefont {P.}~\bibnamefont {M{\"{u}}ller}},
  \bibinfo {author} {\bibfnamefont {W.}~\bibnamefont
  {N{\"{o}}rtersh{\"{a}}user}}, \bibinfo {author} {\bibfnamefont
  {M.}~\bibnamefont {Nunnemann}}, \bibinfo {author} {\bibfnamefont
  {G.}~\bibnamefont {Passler}}, \bibinfo {author} {\bibfnamefont
  {A.}~\bibnamefont {Schmitt}}, \bibinfo {author} {\bibfnamefont
  {N.}~\bibnamefont {Trautmann}},\ and\ \bibinfo {author} {\bibfnamefont
  {A.}~\bibnamefont {Waldek}},\ }\href {https://doi.org/10.1007/s002160051370}
  {\bibfield  {journal} {\bibinfo  {journal} {Fresenius. J. Anal. Chem.}\
  }\textbf {\bibinfo {volume} {364}},\ \bibinfo {pages} {471} (\bibinfo {year}
  {1999})}\BibitemShut {NoStop}%
\bibitem [{\citenamefont {Poulsen}\ and\ \citenamefont
  {Riis}(1988)}]{Poulsen1988}%
  \BibitemOpen
  \bibfield  {author} {\bibinfo {author} {\bibfnamefont {O.}~\bibnamefont
  {Poulsen}}\ and\ \bibinfo {author} {\bibfnamefont {E.}~\bibnamefont {Riis}},\
  }\href {https://doi.org/10.1088/0026-1394/25/3/004} {\bibfield  {journal}
  {\bibinfo  {journal} {Metrologia}\ }\textbf {\bibinfo {volume} {25}},\
  \bibinfo {pages} {147} (\bibinfo {year} {1988})}\BibitemShut {NoStop}%
\bibitem [{\citenamefont {G{\"{o}}tte}\ \emph {et~al.}(2004)\citenamefont
  {G{\"{o}}tte}, \citenamefont {Knaak}, \citenamefont {Kotovski}, \citenamefont
  {Kluge}, \citenamefont {Ewald},\ and\ \citenamefont {Wendt}}]{Gotte2004}%
  \BibitemOpen
  \bibfield  {author} {\bibinfo {author} {\bibfnamefont {S.}~\bibnamefont
  {G{\"{o}}tte}}, \bibinfo {author} {\bibfnamefont {K.-M.}\ \bibnamefont
  {Knaak}}, \bibinfo {author} {\bibfnamefont {N.}~\bibnamefont {Kotovski}},
  \bibinfo {author} {\bibfnamefont {H.-J.}\ \bibnamefont {Kluge}}, \bibinfo
  {author} {\bibfnamefont {G.}~\bibnamefont {Ewald}},\ and\ \bibinfo {author}
  {\bibfnamefont {K.~D.~A.}\ \bibnamefont {Wendt}},\ }\href
  {https://doi.org/10.1063/1.1651635} {\bibfield  {journal} {\bibinfo
  {journal} {Rev. Sci. Instrum.}\ }\textbf {\bibinfo {volume} {75}},\ \bibinfo
  {pages} {1039} (\bibinfo {year} {2004})}\BibitemShut {NoStop}%
\bibitem [{\citenamefont {Krieger}\ \emph {et~al.}(2011)\citenamefont
  {Krieger}, \citenamefont {Geppert}, \citenamefont {Catherall}, \citenamefont
  {Hochschulz}, \citenamefont {Kr{\"{a}}mer}, \citenamefont {Neugart},
  \citenamefont {Rosendahl}, \citenamefont {Schipper}, \citenamefont
  {Siesling}, \citenamefont {Weinheimer}, \citenamefont {Yordanov},\ and\
  \citenamefont {N{\"{o}}rtersh{\"{a}}user}}]{Krieger2011}%
  \BibitemOpen
  \bibfield  {author} {\bibinfo {author} {\bibfnamefont {A.}~\bibnamefont
  {Krieger}}, \bibinfo {author} {\bibfnamefont {C.}~\bibnamefont {Geppert}},
  \bibinfo {author} {\bibfnamefont {R.}~\bibnamefont {Catherall}}, \bibinfo
  {author} {\bibfnamefont {F.}~\bibnamefont {Hochschulz}}, \bibinfo {author}
  {\bibfnamefont {J.}~\bibnamefont {Kr{\"{a}}mer}}, \bibinfo {author}
  {\bibfnamefont {R.}~\bibnamefont {Neugart}}, \bibinfo {author} {\bibfnamefont
  {S.}~\bibnamefont {Rosendahl}}, \bibinfo {author} {\bibfnamefont
  {J.}~\bibnamefont {Schipper}}, \bibinfo {author} {\bibfnamefont
  {E.}~\bibnamefont {Siesling}}, \bibinfo {author} {\bibfnamefont
  {C.}~\bibnamefont {Weinheimer}}, \bibinfo {author} {\bibfnamefont {D.~T.}\
  \bibnamefont {Yordanov}},\ and\ \bibinfo {author} {\bibfnamefont
  {W.}~\bibnamefont {N{\"{o}}rtersh{\"{a}}user}},\ }\href
  {https://doi.org/10.1016/j.nima.2010.12.145} {\bibfield  {journal} {\bibinfo
  {journal} {Nucl. Instruments Methods Phys. Res. Sect. A}\ }\textbf {\bibinfo
  {volume} {632}},\ \bibinfo {pages} {23} (\bibinfo {year} {2011})}\BibitemShut
  {NoStop}%
\bibitem [{\citenamefont {Baczynska}\ \emph {et~al.}(2010)\citenamefont
  {Baczynska}, \citenamefont {Billowes}, \citenamefont {Campbell},
  \citenamefont {Charlwood}, \citenamefont {Cheal}, \citenamefont {Eronen},
  \citenamefont {Forest}, \citenamefont {Jokinen}, \citenamefont {Kessler},
  \citenamefont {Moore}, \citenamefont {R{\"{u}}ffer}, \citenamefont
  {Tungate},\ and\ \citenamefont {{\"{A}}yst{\"{o}}}}]{Baczynska2010}%
  \BibitemOpen
  \bibfield  {author} {\bibinfo {author} {\bibfnamefont {K.}~\bibnamefont
  {Baczynska}}, \bibinfo {author} {\bibfnamefont {J.}~\bibnamefont {Billowes}},
  \bibinfo {author} {\bibfnamefont {P.}~\bibnamefont {Campbell}}, \bibinfo
  {author} {\bibfnamefont {F.~C.}\ \bibnamefont {Charlwood}}, \bibinfo {author}
  {\bibfnamefont {B.}~\bibnamefont {Cheal}}, \bibinfo {author} {\bibfnamefont
  {T.}~\bibnamefont {Eronen}}, \bibinfo {author} {\bibfnamefont {D.~H.}\
  \bibnamefont {Forest}}, \bibinfo {author} {\bibfnamefont {A.}~\bibnamefont
  {Jokinen}}, \bibinfo {author} {\bibfnamefont {T.}~\bibnamefont {Kessler}},
  \bibinfo {author} {\bibfnamefont {I.~D.}\ \bibnamefont {Moore}}, \bibinfo
  {author} {\bibfnamefont {M.}~\bibnamefont {R{\"{u}}ffer}}, \bibinfo {author}
  {\bibfnamefont {G.}~\bibnamefont {Tungate}},\ and\ \bibinfo {author}
  {\bibfnamefont {J.}~\bibnamefont {{\"{A}}yst{\"{o}}}},\ }\href
  {https://doi.org/10.1088/0954-3899/37/10/105103} {\bibfield  {journal}
  {\bibinfo  {journal} {J. Phys. G Nucl. Part. Phys.}\ }\textbf {\bibinfo
  {volume} {37}},\ \bibinfo {pages} {105103} (\bibinfo {year}
  {2010})}\BibitemShut {NoStop}%
\bibitem [{\citenamefont {Dinneen}\ \emph {et~al.}(1991)\citenamefont
  {Dinneen}, \citenamefont {{Berrah Mansour}}, \citenamefont {Kurtz},\ and\
  \citenamefont {Young}}]{Dinneen1991}%
  \BibitemOpen
  \bibfield  {author} {\bibinfo {author} {\bibfnamefont {T.~P.}\ \bibnamefont
  {Dinneen}}, \bibinfo {author} {\bibfnamefont {N.}~\bibnamefont {{Berrah
  Mansour}}}, \bibinfo {author} {\bibfnamefont {C.}~\bibnamefont {Kurtz}},\
  and\ \bibinfo {author} {\bibfnamefont {L.}~\bibnamefont {Young}},\ }\href
  {https://doi.org/10.1103/PhysRevA.43.4824} {\bibfield  {journal} {\bibinfo
  {journal} {Phys. Rev. A}\ }\textbf {\bibinfo {volume} {43}},\ \bibinfo
  {pages} {4824} (\bibinfo {year} {1991})}\BibitemShut {NoStop}%
\bibitem [{\citenamefont {Th{\"{u}}mmler}\ \emph {et~al.}(2009)\citenamefont
  {Th{\"{u}}mmler}, \citenamefont {Marx},\ and\ \citenamefont
  {Weinheimer}}]{Thummler2009}%
  \BibitemOpen
  \bibfield  {author} {\bibinfo {author} {\bibfnamefont {T.}~\bibnamefont
  {Th{\"{u}}mmler}}, \bibinfo {author} {\bibfnamefont {R.}~\bibnamefont
  {Marx}},\ and\ \bibinfo {author} {\bibfnamefont {C.}~\bibnamefont
  {Weinheimer}},\ }\href {https://doi.org/10.1088/1367-2630/11/10/103007}
  {\bibfield  {journal} {\bibinfo  {journal} {New J. Phys.}\ }\textbf {\bibinfo
  {volume} {11}},\ \bibinfo {pages} {103007} (\bibinfo {year}
  {2009})}\BibitemShut {NoStop}%
\bibitem [{\citenamefont {Rest}\ \emph {et~al.}(2019)\citenamefont {Rest},
  \citenamefont {Winzen}, \citenamefont {Bauer}, \citenamefont {Berendes},
  \citenamefont {Meisner}, \citenamefont {Th{\"{u}}mmler}, \citenamefont
  {W{\"{u}}stling},\ and\ \citenamefont {Weinheimer}}]{Rest2019}%
  \BibitemOpen
  \bibfield  {author} {\bibinfo {author} {\bibfnamefont {O.}~\bibnamefont
  {Rest}}, \bibinfo {author} {\bibfnamefont {D.}~\bibnamefont {Winzen}},
  \bibinfo {author} {\bibfnamefont {S.}~\bibnamefont {Bauer}}, \bibinfo
  {author} {\bibfnamefont {R.}~\bibnamefont {Berendes}}, \bibinfo {author}
  {\bibfnamefont {J.}~\bibnamefont {Meisner}}, \bibinfo {author} {\bibfnamefont
  {T.}~\bibnamefont {Th{\"{u}}mmler}}, \bibinfo {author} {\bibfnamefont
  {S.}~\bibnamefont {W{\"{u}}stling}},\ and\ \bibinfo {author} {\bibfnamefont
  {C.}~\bibnamefont {Weinheimer}},\ }\href
  {https://doi.org/10.1088/1681-7575/ab2997} {\bibfield  {journal} {\bibinfo
  {journal} {Metrologia}\ }\textbf {\bibinfo {volume} {56}},\ \bibinfo {pages}
  {045007} (\bibinfo {year} {2019})}\BibitemShut {NoStop}%
\bibitem [{\citenamefont {Aker}\ \emph {et~al.}(2019)\citenamefont {Aker},
  \citenamefont {Altenm{\"{u}}ller}, \citenamefont {Arenz}, \citenamefont
  {Babutzka}, \citenamefont {Barrett}, \citenamefont {Bauer}, \citenamefont
  {Beck}, \citenamefont {Beglarian}, \citenamefont {Behrens}, \citenamefont
  {Bergmann}, \citenamefont {Besserer}, \citenamefont {Blaum}, \citenamefont
  {Block}, \citenamefont {Bobien}, \citenamefont {Bokeloh} \emph
  {et~al.}}]{Aker2019}%
  \BibitemOpen
  \bibfield  {author} {\bibinfo {author} {\bibfnamefont {M.}~\bibnamefont
  {Aker}}, \bibinfo {author} {\bibfnamefont {K.}~\bibnamefont
  {Altenm{\"{u}}ller}}, \bibinfo {author} {\bibfnamefont {M.}~\bibnamefont
  {Arenz}}, \bibinfo {author} {\bibfnamefont {M.}~\bibnamefont {Babutzka}},
  \bibinfo {author} {\bibfnamefont {J.}~\bibnamefont {Barrett}}, \bibinfo
  {author} {\bibfnamefont {S.}~\bibnamefont {Bauer}}, \bibinfo {author}
  {\bibfnamefont {M.}~\bibnamefont {Beck}}, \bibinfo {author} {\bibfnamefont
  {A.}~\bibnamefont {Beglarian}}, \bibinfo {author} {\bibfnamefont
  {J.}~\bibnamefont {Behrens}}, \bibinfo {author} {\bibfnamefont
  {T.}~\bibnamefont {Bergmann}}, \bibinfo {author} {\bibfnamefont
  {U.}~\bibnamefont {Besserer}}, \bibinfo {author} {\bibfnamefont
  {K.}~\bibnamefont {Blaum}}, \bibinfo {author} {\bibfnamefont
  {F.}~\bibnamefont {Block}}, \bibinfo {author} {\bibfnamefont
  {S.}~\bibnamefont {Bobien}}, \bibinfo {author} {\bibfnamefont
  {K.}~\bibnamefont {Bokeloh}}, \emph {et~al.},\ }\href
  {https://doi.org/10.1103/PhysRevLett.123.221802} {\bibfield  {journal}
  {\bibinfo  {journal} {Phys. Rev. Lett.}\ }\textbf {\bibinfo {volume} {123}},\
  \bibinfo {pages} {221802(10)} (\bibinfo {year} {2019})}\BibitemShut {NoStop}%
\bibitem [{\citenamefont {Kr{\"{a}}mer}\ \emph {et~al.}(2018)\citenamefont
  {Kr{\"{a}}mer}, \citenamefont {K{\"{o}}nig}, \citenamefont {Geppert},
  \citenamefont {Imgram}, \citenamefont {Maa{\ss}}, \citenamefont {Meisner},
  \citenamefont {Otten}, \citenamefont {Passon}, \citenamefont {Ratajczyk},
  \citenamefont {Ullmann},\ and\ \citenamefont
  {N{\"{o}}rtersh{\"{a}}user}}]{Kramer2018}%
  \BibitemOpen
  \bibfield  {author} {\bibinfo {author} {\bibfnamefont {J.}~\bibnamefont
  {Kr{\"{a}}mer}}, \bibinfo {author} {\bibfnamefont {K.}~\bibnamefont
  {K{\"{o}}nig}}, \bibinfo {author} {\bibfnamefont {C.}~\bibnamefont
  {Geppert}}, \bibinfo {author} {\bibfnamefont {P.}~\bibnamefont {Imgram}},
  \bibinfo {author} {\bibfnamefont {B.}~\bibnamefont {Maa{\ss}}}, \bibinfo
  {author} {\bibfnamefont {J.}~\bibnamefont {Meisner}}, \bibinfo {author}
  {\bibfnamefont {E.~W.}\ \bibnamefont {Otten}}, \bibinfo {author}
  {\bibfnamefont {S.}~\bibnamefont {Passon}}, \bibinfo {author} {\bibfnamefont
  {T.}~\bibnamefont {Ratajczyk}}, \bibinfo {author} {\bibfnamefont
  {J.}~\bibnamefont {Ullmann}},\ and\ \bibinfo {author} {\bibfnamefont
  {W.}~\bibnamefont {N{\"{o}}rtersh{\"{a}}user}},\ }\href
  {https://doi.org/10.1088/1681-7575/aaabe0} {\bibfield  {journal} {\bibinfo
  {journal} {Metrologia}\ }\textbf {\bibinfo {volume} {55}},\ \bibinfo {pages}
  {268} (\bibinfo {year} {2018})}\BibitemShut {NoStop}%
\bibitem [{\citenamefont {Barton}\ \emph {et~al.}(2000)\citenamefont {Barton},
  \citenamefont {Donald}, \citenamefont {Lucas}, \citenamefont {Stevens},
  \citenamefont {Steane},\ and\ \citenamefont {Stacey}}]{Barton2000}%
  \BibitemOpen
  \bibfield  {author} {\bibinfo {author} {\bibfnamefont {P.~A.}\ \bibnamefont
  {Barton}}, \bibinfo {author} {\bibfnamefont {C.~J.~S.}\ \bibnamefont
  {Donald}}, \bibinfo {author} {\bibfnamefont {D.~M.}\ \bibnamefont {Lucas}},
  \bibinfo {author} {\bibfnamefont {D.~A.}\ \bibnamefont {Stevens}}, \bibinfo
  {author} {\bibfnamefont {A.~M.}\ \bibnamefont {Steane}},\ and\ \bibinfo
  {author} {\bibfnamefont {D.~N.}\ \bibnamefont {Stacey}},\ }\href
  {https://journals.aps.org/pra/pdf/10.1103/PhysRevA.62.032503} {\bibfield
  {journal} {\bibinfo  {journal} {Phys. Rev. A}\ }\textbf {\bibinfo {volume}
  {62}},\ \bibinfo {pages} {032503} (\bibinfo {year} {2000})}\BibitemShut
  {NoStop}%
\bibitem [{\citenamefont {Cohen‐Tannoudji}\ \emph {et~al.}(2008)\citenamefont
  {Cohen‐Tannoudji}, \citenamefont {Dupont‐Roc},\ and\ \citenamefont
  {Grynberg}}]{CohenTannoudji2008}%
  \BibitemOpen
  \bibfield  {author} {\bibinfo {author} {\bibfnamefont {C.}~\bibnamefont
  {Cohen‐Tannoudji}}, \bibinfo {author} {\bibfnamefont {J.}~\bibnamefont
  {Dupont‐Roc}},\ and\ \bibinfo {author} {\bibfnamefont {G.}~\bibnamefont
  {Grynberg}},\ }\href {https://doi.org/10.1002/9783527617197} {\emph {\bibinfo
  {title} {Atom-Photon Interactions}}}\ (\bibinfo  {publisher} {Wiley-VCH},\
  \bibinfo {year} {2008})\BibitemShut {NoStop}%
\bibitem [{\citenamefont {Dalibard}\ and\ \citenamefont
  {Cohen-Tannoudji}(1985)}]{Dalibard1985}%
  \BibitemOpen
  \bibfield  {author} {\bibinfo {author} {\bibfnamefont {J.}~\bibnamefont
  {Dalibard}}\ and\ \bibinfo {author} {\bibfnamefont {C.}~\bibnamefont
  {Cohen-Tannoudji}},\ }\href {https://doi.org/10.1088/0022-3700/18/8/019}
  {\bibfield  {journal} {\bibinfo  {journal} {J. Phys. B At. Mol. Phys}\
  }\textbf {\bibinfo {volume} {18}},\ \bibinfo {pages} {1661} (\bibinfo {year}
  {1985})}\BibitemShut {NoStop}%
\bibitem [{\citenamefont {Marksteiner}\ \emph {et~al.}(1995)\citenamefont
  {Marksteiner}, \citenamefont {Walser}, \citenamefont {Marte},\ and\
  \citenamefont {Zoller}}]{Marksteiner1995}%
  \BibitemOpen
  \bibfield  {author} {\bibinfo {author} {\bibfnamefont {S.}~\bibnamefont
  {Marksteiner}}, \bibinfo {author} {\bibfnamefont {R.}~\bibnamefont {Walser}},
  \bibinfo {author} {\bibfnamefont {P.}~\bibnamefont {Marte}},\ and\ \bibinfo
  {author} {\bibfnamefont {P.}~\bibnamefont {Zoller}},\ }\href
  {https://doi.org/10.1007/BF01135856} {\bibfield  {journal} {\bibinfo
  {journal} {Appl. Phys. B}\ }\textbf {\bibinfo {volume} {60}},\ \bibinfo
  {pages} {145} (\bibinfo {year} {1995})}\BibitemShut {NoStop}%
\bibitem [{\citenamefont {Walser}\ \emph {et~al.}(1994)\citenamefont {Walser},
  \citenamefont {Cooper},\ and\ \citenamefont {Zoller}}]{Walser1994}%
  \BibitemOpen
  \bibfield  {author} {\bibinfo {author} {\bibfnamefont {R.}~\bibnamefont
  {Walser}}, \bibinfo {author} {\bibfnamefont {J.}~\bibnamefont {Cooper}},\
  and\ \bibinfo {author} {\bibfnamefont {P.}~\bibnamefont {Zoller}},\ }\href
  {https://doi.org/10.1103/PhysRevA.50.4303} {\bibfield  {journal} {\bibinfo
  {journal} {Phys. Rev. A}\ }\textbf {\bibinfo {volume} {50}},\ \bibinfo
  {pages} {4303} (\bibinfo {year} {1994})}\BibitemShut {NoStop}%
\bibitem [{\citenamefont {Rosenbluh}\ \emph {et~al.}(1998)\citenamefont
  {Rosenbluh}, \citenamefont {Rosenhouse-Dantsker}, \citenamefont
  {Wilson-Gordon}, \citenamefont {Levenson},\ and\ \citenamefont
  {Walser}}]{Rosenbluh1998}%
  \BibitemOpen
  \bibfield  {author} {\bibinfo {author} {\bibfnamefont {M.}~\bibnamefont
  {Rosenbluh}}, \bibinfo {author} {\bibfnamefont {A.}~\bibnamefont
  {Rosenhouse-Dantsker}}, \bibinfo {author} {\bibfnamefont {A.}~\bibnamefont
  {Wilson-Gordon}}, \bibinfo {author} {\bibfnamefont {M.}~\bibnamefont
  {Levenson}},\ and\ \bibinfo {author} {\bibfnamefont {R.}~\bibnamefont
  {Walser}},\ }\href {https://doi.org/10.1016/S0030-4018(97)00485-9} {\bibfield
   {journal} {\bibinfo  {journal} {Opt. Commun.}\ }\textbf {\bibinfo {volume}
  {146}},\ \bibinfo {pages} {158} (\bibinfo {year} {1998})}\BibitemShut
  {NoStop}%
\bibitem [{\citenamefont {McIntyre}\ \emph {et~al.}(1993)\citenamefont
  {McIntyre}, \citenamefont {Fairchild}, \citenamefont {Cooper},\ and\
  \citenamefont {Walser}}]{McIntyre1993}%
  \BibitemOpen
  \bibfield  {author} {\bibinfo {author} {\bibfnamefont {D.~H.}\ \bibnamefont
  {McIntyre}}, \bibinfo {author} {\bibfnamefont {C.~E.}\ \bibnamefont
  {Fairchild}}, \bibinfo {author} {\bibfnamefont {J.}~\bibnamefont {Cooper}},\
  and\ \bibinfo {author} {\bibfnamefont {R.}~\bibnamefont {Walser}},\ }\href
  {https://doi.org/10.1364/ol.18.001816} {\bibfield  {journal} {\bibinfo
  {journal} {Opt. Lett.}\ }\textbf {\bibinfo {volume} {18}},\ \bibinfo {pages}
  {1816} (\bibinfo {year} {1993})}\BibitemShut {NoStop}%
\bibitem [{\citenamefont {Sturm}\ \emph {et~al.}(2014)\citenamefont {Sturm},
  \citenamefont {Rein}, \citenamefont {Walther},\ and\ \citenamefont
  {Walser}}]{Sturm2014}%
  \BibitemOpen
  \bibfield  {author} {\bibinfo {author} {\bibfnamefont {M.~R.}\ \bibnamefont
  {Sturm}}, \bibinfo {author} {\bibfnamefont {B.}~\bibnamefont {Rein}},
  \bibinfo {author} {\bibfnamefont {T.}~\bibnamefont {Walther}},\ and\ \bibinfo
  {author} {\bibfnamefont {R.}~\bibnamefont {Walser}},\ }\href
  {https://doi.org/10.1364/josab.31.001964} {\bibfield  {journal} {\bibinfo
  {journal} {J. Opt. Soc. Am. B}\ }\textbf {\bibinfo {volume} {31}},\ \bibinfo
  {pages} {1964} (\bibinfo {year} {2014})}\BibitemShut {NoStop}%
\bibitem [{\citenamefont {Brion}\ \emph {et~al.}(2007)\citenamefont {Brion},
  \citenamefont {Pedersen},\ and\ \citenamefont {M{\o}lmer}}]{Brion2007}%
  \BibitemOpen
  \bibfield  {author} {\bibinfo {author} {\bibfnamefont {E.}~\bibnamefont
  {Brion}}, \bibinfo {author} {\bibfnamefont {L.~H.}\ \bibnamefont
  {Pedersen}},\ and\ \bibinfo {author} {\bibfnamefont {K.}~\bibnamefont
  {M{\o}lmer}},\ }\href {https://doi.org/10.1088/1751-8113/40/5/011} {\bibfield
   {journal} {\bibinfo  {journal} {J. Phys. A Math. Theor. J.}\ }\textbf
  {\bibinfo {volume} {40}},\ \bibinfo {pages} {1033} (\bibinfo {year}
  {2007})}\BibitemShut {NoStop}%
\bibitem [{\citenamefont {Abramowitz}\ and\ \citenamefont
  {Stegun}(1964)}]{Abramowitz1964}%
  \BibitemOpen
  \bibfield  {author} {\bibinfo {author} {\bibfnamefont {M.}~\bibnamefont
  {Abramowitz}}\ and\ \bibinfo {author} {\bibfnamefont {I.~A.}\ \bibnamefont
  {Stegun}},\ }\href {https://doi.org/10.1115/1.3625776} {\emph {\bibinfo
  {title} {{Handbook of Mathematical Functions With Formulas, Graphs, and
  Mathematical Tables}}}},\ \bibinfo {edition} {10th}\ ed.\ (\bibinfo
  {publisher} {US Government Printing Office},\ \bibinfo {address}
  {Washington},\ \bibinfo {year} {1964})\BibitemShut {NoStop}%
\bibitem [{\citenamefont {Wang}\ \emph {et~al.}(2017)\citenamefont {Wang},
  \citenamefont {Audi}, \citenamefont {Kondev}, \citenamefont {Huang},
  \citenamefont {Naimi},\ and\ \citenamefont {Xu}}]{Wang2017}%
  \BibitemOpen
  \bibfield  {author} {\bibinfo {author} {\bibfnamefont {M.}~\bibnamefont
  {Wang}}, \bibinfo {author} {\bibfnamefont {G.}~\bibnamefont {Audi}}, \bibinfo
  {author} {\bibfnamefont {F.~G.}\ \bibnamefont {Kondev}}, \bibinfo {author}
  {\bibfnamefont {W.~J.}\ \bibnamefont {Huang}}, \bibinfo {author}
  {\bibfnamefont {S.}~\bibnamefont {Naimi}},\ and\ \bibinfo {author}
  {\bibfnamefont {X.}~\bibnamefont {Xu}},\ }\href
  {https://doi.org/10.1088/1674-1137/41/3/030003} {\bibfield  {journal}
  {\bibinfo  {journal} {Chinese Phys. C}\ }\textbf {\bibinfo {volume} {41}},\
  \bibinfo {pages} {030003} (\bibinfo {year} {2017})}\BibitemShut {NoStop}%
\bibitem [{\citenamefont {Kramida}\ \emph {et~al.}(2018)\citenamefont
  {Kramida}, \citenamefont {Ralchenko}, \citenamefont {Reader},\ and\
  \citenamefont {Team}}]{NIST18}%
  \BibitemOpen
  \bibfield  {author} {\bibinfo {author} {\bibfnamefont {A.}~\bibnamefont
  {Kramida}}, \bibinfo {author} {\bibfnamefont {Y.}~\bibnamefont {Ralchenko}},
  \bibinfo {author} {\bibfnamefont {J.}~\bibnamefont {Reader}},\ and\ \bibinfo
  {author} {\bibfnamefont {N.~A.}\ \bibnamefont {Team}},\ }\href
  {https://physics.nist.gov/asd} {\bibinfo {title} {{NIST Atomic Spectra
  Database (ver. 5.6.1)}}} (\bibinfo {year} {2018})\BibitemShut {NoStop}%
\bibitem [{\citenamefont {Shi}\ \emph {et~al.}(2017)\citenamefont {Shi},
  \citenamefont {Gebert}, \citenamefont {Gorges}, \citenamefont {Kaufmann},
  \citenamefont {N{\"{o}}rtersh{\"{a}}user}, \citenamefont {Sahoo},
  \citenamefont {Surzhykov}, \citenamefont {Yerokhin}, \citenamefont
  {Berengut}, \citenamefont {Wolf}, \citenamefont {Heip},\ and\ \citenamefont
  {Schmidt}}]{Shi2017}%
  \BibitemOpen
  \bibfield  {author} {\bibinfo {author} {\bibfnamefont {C.}~\bibnamefont
  {Shi}}, \bibinfo {author} {\bibfnamefont {F.}~\bibnamefont {Gebert}},
  \bibinfo {author} {\bibfnamefont {C.}~\bibnamefont {Gorges}}, \bibinfo
  {author} {\bibfnamefont {S.}~\bibnamefont {Kaufmann}}, \bibinfo {author}
  {\bibfnamefont {W.}~\bibnamefont {N{\"{o}}rtersh{\"{a}}user}}, \bibinfo
  {author} {\bibfnamefont {B.~K.}\ \bibnamefont {Sahoo}}, \bibinfo {author}
  {\bibfnamefont {A.}~\bibnamefont {Surzhykov}}, \bibinfo {author}
  {\bibfnamefont {V.~A.}\ \bibnamefont {Yerokhin}}, \bibinfo {author}
  {\bibfnamefont {J.~C.}\ \bibnamefont {Berengut}}, \bibinfo {author}
  {\bibfnamefont {F.}~\bibnamefont {Wolf}}, \bibinfo {author} {\bibfnamefont
  {J.~C.}\ \bibnamefont {Heip}},\ and\ \bibinfo {author} {\bibfnamefont
  {P.~O.}\ \bibnamefont {Schmidt}},\ }\href
  {https://doi.org/10.1007/s00340-016-6572-z} {\bibfield  {journal} {\bibinfo
  {journal} {Appl. Phys. B}\ }\textbf {\bibinfo {volume} {123}},\ \bibinfo
  {pages} {2} (\bibinfo {year} {2017})}\BibitemShut {NoStop}%
\bibitem [{\citenamefont {Guan}\ \emph {et~al.}(2015)\citenamefont {Guan},
  \citenamefont {Huang}, \citenamefont {Liu}, \citenamefont {Bian},
  \citenamefont {Shao},\ and\ \citenamefont {Gao}}]{Guan2015}%
  \BibitemOpen
  \bibfield  {author} {\bibinfo {author} {\bibfnamefont {H.}~\bibnamefont
  {Guan}}, \bibinfo {author} {\bibfnamefont {Y.}~\bibnamefont {Huang}},
  \bibinfo {author} {\bibfnamefont {P.~L.}\ \bibnamefont {Liu}}, \bibinfo
  {author} {\bibfnamefont {W.}~\bibnamefont {Bian}}, \bibinfo {author}
  {\bibfnamefont {H.}~\bibnamefont {Shao}},\ and\ \bibinfo {author}
  {\bibfnamefont {K.~L.}\ \bibnamefont {Gao}},\ }\href
  {https://doi.org/10.1088/1674-1056/24/5/054213} {\bibfield  {journal}
  {\bibinfo  {journal} {Chinese Phys. B}\ }\textbf {\bibinfo {volume} {24}},\
  \bibinfo {pages} {054213} (\bibinfo {year} {2015})}\BibitemShut {NoStop}%
\bibitem [{\citenamefont {Shore}(1990)}]{ShoreChap}%
  \BibitemOpen
  \bibfield  {author} {\bibinfo {author} {\bibfnamefont {B.~W.}\ \bibnamefont
  {Shore}},\ }\href@noop {} {\emph {\bibinfo {title} {{The theory of coherent
  atomic excitation}}}},\ Vol.\ \bibinfo {volume} {1 and 2, Chap. 2.6, 20.7,
  20.8}\ (\bibinfo  {publisher} {Wiley-Interscience},\ \bibinfo {year}
  {1990})\BibitemShut {NoStop}%
\end{thebibliography}%
\end{document}